\documentclass[journal,twoside,web]{ieeecolor}

\usepackage{generic}
\usepackage{amsmath,amssymb,amsfonts}
\usepackage{hyperref}
\usepackage{textcomp}

\def\BibTeX{{\rm B\kern-.05em{\sc i\kern-.025em b}\kern-.08em
    T\kern-.1667em\lower.7ex\hbox{E}\kern-.125emX}}
    
\usepackage{bm}
\usepackage{balance}
\usepackage{microtype}
\usepackage{hyperref}
\usepackage{amsmath}
\usepackage{mathtools}
\usepackage{amssymb}
\usepackage{balance}
\usepackage{comment}
\usepackage[pdftex]{graphicx}
\usepackage{booktabs} 
\usepackage{marvosym}
\usepackage{pdflscape}
\usepackage{multirow}
\usepackage{nicefrac}
\usepackage{tabu}
\usepackage{pdflscape}
\usepackage{lipsum}
\usepackage{float}
\usepackage{stfloats}

\usepackage{bm}
\renewcommand{\vec}[1]{\bm{{#1}}}

\usepackage{lscape}
\usepackage{longtable}

\usepackage{color, colortbl}

\definecolor{white}{rgb}    {1.00, 1.00, 1.00} 
\definecolor{lightgray}{rgb}{0.73, 0.73, 0.73} 
\definecolor{orange}{rgb}   {0.93, 0.47, 0.20} 
\definecolor{green}{rgb}    {0.00, 0.60, 0.53} 
\definecolor{red}{rgb}      {0.80, 0.20, 0.07} 
\definecolor{blue}{rgb}     {0.00, 0.47, 0.73} 
\definecolor{pink}{rgb}     {0.93, 0.20, 0.47} 
\definecolor{lightblue}{rgb}{0.20, 0.73, 0.93} 

\usepackage[linesnumbered,ruled,vlined]{algorithm2e}
\SetKwInput{KwInput}{Input}
\SetKwInput{KwOutput}{Output}

\usepackage{tocloft}
\usepackage[english]{babel}
\usepackage{csquotes}
\addto\captionsenglish{}
\newcommand{\nocontentsline}[3]{}
\newcommand{\tocless}[2]{\bgroup\let\addcontentsline=\nocontentsline#1{#2}\egroup}

\usepackage[backend=biber, style=ieee, sorting=none, url=false, doi=false, eprint=false, maxbibnames=6]{biblatex}
\DeclareFieldFormat{labelnumberwidth}{\mkbibbrackets{#1}}

\begin{document}
\title{A Deep Generative Model for Five-Class Sleep Staging with Arbitrary Sensor Input}

\author{Hans van Gorp, \IEEEmembership{Student Member, IEEE}, Merel M. van Gilst, Pedro Fonseca, Fokke B. van Meulen, Johannes P. van Dijk, \IEEEmembership{Senior Member, IEEE}, Sebastiaan Overeem, and Ruud J.G. van Sloun, \IEEEmembership{Senior Member, IEEE}
\thanks{This work was performed within the IMPULSE framework of the Eindhoven MedTech Innovation Center (e/MTIC, incorporating Eindhoven University of Technology, Philips Research, and Sleep Medicine Center, Kempenhaeghe Foundation), including a PPS supplement from the Dutch Ministry of Economic Affairs and Climate Policy.}
\thanks{Hans van Gorp, Merel M. van Gilst, Pedro Fonseca, Fokke B. van Meulen, Johannes P. van Dijk, Sebastiaan Overeem, and Ruud J.G. van Sloun are with the Department of Electrical Engineering, Eindhoven University of Technology, 5612AZ Eindhoven, The Netherlands (e-mail: h.v.gorp@tue.nl, m.m.v.gilst@tue.nl, f.b.v.meulen@tue.nl, j.p.v.dijk@tue.nl, s.overeem@tue.nl, r.j.g.v.sloun@tue.nl).}
\thanks{Hans van Gorp and Pedro Fonseca are with Philips Sleep and Respiratory Care, 5656AE Eindhoven, The Netherlands (e-mail: pedro.fonseca@philips.com).}
\thanks{Merel M. van Gilst, Fokke B. van Meulen, Johannes P. van Dijk, and Sebastiaan Overeem are with Sleep Medicine Centre Kempenhaeghe, 5591VE Heeze, The Netherlands.}
\thanks{Johannes P. van Dijk is with the Department of Orthodontics, Ulm University, 89081 Ulm, Germany.}
\thanks{Supplemental information available at:\\ \url{https://github.com/HansvanGorp/FSDM-supplement}}
\vspace{-1em}
}

\maketitle
\begin{abstract}
Gold-standard sleep scoring is based on epoch-based assignment of sleep stages based on a combination of EEG, EOG and EMG signals.  However, a polysomnographic recording  consists of many other signals that could  be used for sleep staging, including cardio-respiratory modalities. Leveraging this signal variety would offer important advantages, for example increasing reliability, resilience to signal loss, and application to long-term non-obtrusive recordings. We developed a deep generative model for automatic sleep staging from a plurality of sensors and any -arbitrary- combination thereof. We trained a score-based diffusion model using a dataset of 1947 expert-labelled overnight recordings with 36 different signals, and achieved zero-shot inference on any sensor set by leveraging a novel Bayesian factorization of the score function across the sensors. On single-channel EEG, the model reaches the performance limit in terms of polysomnography inter-rater agreement (5- class accuracy 85.6\%, Cohen’s kappa 0.791). Moreover, the method offers full flexibility to use any sensor set, for example finger photoplethysmography, nasal flow and thoracic respiratory movements, (5-class accuracy 79.0\%, Cohen’s kappa of 0.697), or even derivations very unconventional for sleep staging, such as tibialis and sternocleidomastoid EMG (5-class accuracy 71.0\%, kappa 0.575). Additionally, we propose a novel interpretability metric in terms of information gain per sensor and show this is linearly correlated with classification performance. Finally, our model allows for post- hoc addition of entirely new sensor modalities by merely training a score estimator on the novel input instead of having to retrain from scratch on all inputs.  
\end{abstract}

\begin{IEEEkeywords}
Automatic Sleep Staging, Deep Learning, Generative AI, Diffusion Models, Score-based Diffusion Models.
\end{IEEEkeywords}

\tocless\section{Introduction} \noindent
Sleep stage scoring is an essential tool in the clinical assessment of sleep and the diagnosis of sleep disorders. Traditionally, sleep staging has relied on overnight polysomnographic (PSG) recordings which at least include electroencephalography (EEG), electrooculography (EOG) and electromyography (EMG). The accepted gold standard is for experienced human scorers to perform this sleep staging manually, following the guidelines of the American Academy of Sleep Medicine (AASM) \cite{AASM}. Accordingly, each 30-second segment of sleep, known as an epoch, is scored as belonging to one of five stages: Wake (W), Rapid Eye Movement (REM), or non-REM (NREM) stages 1-3 based on the visual recognition of established patterns on EEG, EOG and EMG signals. The representation of a sequence of sleep stages over the night is called a hypnogram. The visual analysis of its characteristics, such as the distribution and continuity of sleep stages help drive clinical interpretation \cite{Woerd2023}.

There are several challenges in sleep scoring, namely its costs, time requirements, and the need for trained personnel. To overcome this, automatic sleep scoring based on PSG has been extensively described in literature. The EEG signal in particular provides a strong basis for automatic sleep staging. A single EEG derivation is often enough to reach performance on par with the human inter-rater agreement \cite{EEG_Staging_review}. While this alleviates some of the costs of human scoring, the EEG needs to be placed above the hairline which can cause patient discomfort, and due to its vulnerability to environmental noise and motion artifacts, is less well-suited for ambulatory or prolonged measurements.

To provide an alternative, the measurement and analysis of surrogate signals has been studied. Surrogate measurements make use of indirect observations, such as movements often associated with wakefulness, or expressions of the sleep stages in autonomic nervous system activity, for example via cardiac and respiratory sensors. Surrogate modalities often described in literature include actigraphy, cardiac activity, respiratory effort, and respiratory flow \cite{SleepV3, CReSS, Mattress_Review}. Because there are no visual sleep scoring rules for these signals, analysis must be performed automatically. Most successful approaches rely on machine learning techniques on measurements of one or more signals, using as training ground-truth sleep stages derived from a human-scored, simultaneously recorded PSG study. Many approaches address a simplified sleep staging set-up, distinguishing only between sleep and wake, or distinguishing between 4 classes instead of the usual 5, where the N1 and N2 classes are merged into a joint N1/N2 class \cite{SleepV3, CReSS, Mattress_Review, cerina2023sleep}.

An unsolved problem remains: between different individual recordings, and more importantly, between measurement setups, the combination of available input signals can vary widely. This can be because of different devices, measurement protocols, sensors inadvertently being disconnected during the recording, or due to interference, noise, and artifacts. Some models described in the literature partially solve this issue and can perform sleep staging on a range of input signals. For example, U-Sleep has been trained specifically to work with any combination of single-channel EEG and single-channel EOG signals, even when using derivations between electrodes not recommended by the AASM \cite{USleep_Perslev}, \cite{resilient_u_sleep}. 

However, no proposed system can easily scale up to new sensors after training on a specific set. Scaling up requires retraining the entire system and collecting numerous new recordings with both old and new signals measured simultaneously. Given the substantial training data required by deep learning systems, successive rounds of data collection pose a practical obstacle to introducing new sensors in practice.

We introduce a deep generative model to accurately and scalably perform sleep staging using any combination of signal modalities used to measure sleep, both PSG-derived as well as surrogate. Such a model can be highly beneficial in clinical practice, as it has the flexibility to adapt to any circumstance and recording protocol, while at the same time being robust to disconnected sensors, missing measurements, and noisy sensors. Our main contributions are as follows:
\begin{itemize}
\item We introduce a novel algorithm based on a Bayesian factorization of score-based diffusion models, which we term Factorized Score-based Diffusion Modeling (FSDM).
\item We show how FSDM can be leveraged to perform joint (zero-shot) inference on arbitrary combinations of input signals while never being trained on them; training only happens on one signal at a time.
\item We provide a natural means of expressing the information gain from each signal in the FSDM framework.
\item We showcase the performance of FSDM on an extensive list of 36 individual signals and their combinations, including improved performance in 5-class sleep staging with several cardio-respiratory modalities.
\end{itemize}
 \tocless\section{Methods}

\tocless\subsection{Factorized score-based diffusion modeling} \noindent
In the proposed framework, individual networks are trained separately on each signal modality. Only during deployment are the models combined into a joint posterior, which allows zero-shot inference on subjects with arbitrary combinations of measurement modalities. While the term `zero shot' on the output side traditionally refers to unseen classes, we use the term in this manuscript  to refer to unseen combinations of input signals,  i.e., a combination of signals that was not seen during training, as the signals were only trained on separately.

To perform this zero-shot inference using arbitrary combinations of sensors, we use a score-based diffusion model as our backbone \cite{conditional_diffusion_2}. This requires an estimate of the posterior score. Our key result is that the posterior score is well estimated by: 

\noindent \begin{multline}
\label{eq:FSDM_rule}
    \underbrace{\nabla_{\vec{y}}\log p\left(\vec{y}|X^{(1:N)}\right)}_{\text{posterior}} \approx \underbrace{\nabla_{\vec{y}}\log q_{\theta^{(0)}}(\vec{y})}_{\text{global prior}} + \frac{1}{N}\sum_{i=1}^{N} \\ \left( \underbrace{\nabla_{\vec{y}}\log q_{\theta^{(i)}}\left(\vec{y}|\vec{x}^{(i)}\right)}_{\text{individual likelihood}} - \underbrace{\nabla_{\vec{y}}\log q_{\theta^{(i)}}(\vec{y})}_{\text{individual prior}} \right),
\end{multline}
where we have factorized the posterior score into its Bayesian components. In equation (\ref{eq:FSDM_rule}), $\vec{y}$ denotes the hypnogram, $X^{(1:N)}$ denotes the combination of input data coming from $N$ different sensors, $\nabla_. \log p(.)$ denotes a true score, and $\nabla_. \log q_{\theta^{(i)}}(.)$ denotes a score as estimated by a neural network with parameters $\theta^{(i)}$, which we will simply call a score-network for the sake of brevity. Each score-network is specific to an input signal with index $i$, for example, $i=1$ could denote a respiratory effort signal and $i=2$ a cardiac signal. 

A crucial insight from equation (\ref{eq:FSDM_rule}) is that the posterior score can be inferred a-posteriori from individually learned likelihood and prior scores. Each of these individual scores is estimated by a score-network trained solely on its own sensor data using denoising score matching techniques. The score-networks are agnostic to the existence of other types of measurement data. The rest of this subsection will now provide a full derivation of the FSDM algorithm. We will use the following notation: \vspace{-3pt}
\begin{itemize}
    \item Let $\vec{y}\in\mathcal{R}^{5\times E}$ be a hypnodensity, i.e., the sleep stage probabilities of size 5 stages by number of epochs $E$.
    \item A hypnogram $\vec{h}\in [W,N1,N2,N3,R]^E$ can be expressed as a hypnodensity $\vec{y}$ through one-hot encoding.
    \item Let $\vec{x}\in \mathcal{R}^{E \cdot F \cdot 30}$ be a signal measured concurrently with the hypnogram at sampling frequency $F$.
    \item A collection of input signals can be written as: \\ $X^{(1:N)} = [\vec{x}^{(1)},\vec{x}^{(2)},\dots,\vec{x}^{(N)}]$.
\end{itemize}

 \subsubsection{Factorized posterior score} \noindent
We are interested in estimating the probability distribution of hypnograms given a set of measurements signals, expressed as $p\left(\vec{y}|X^{(1:N)}\right)$. This estimation problem can be re-written using Bayes' rule as:
\begin{equation}
    p\left(\vec{y}|X^{(1:N)}\right) = \frac{ p(\vec{y})}{p\left(X^{(1:N)}\right)}p\left(X^{(1:N)}|\vec{y}\right).
\end{equation}
By assuming that the individual input signals $\vec{x}^{(i)}$ are conditionally independent given $\vec{y}$, we arrive at the Naive Bayes estimator:

\begin{equation}
    p\left(\vec{y}|X^{(1:N)}\right) = \frac{ p(\vec{y})}{p\left(X^{(1:N)}\right)}\prod_{i=1}^{N}p\left(x^{(i)}|\vec{y}\right).
\end{equation}
Bayes' rule can then be applied a second time, but now to the individual likelihood terms, to arrive at:
\begin{equation}
\label{eq:combination_with_normalization}
    p\left(\vec{y}|X^{(1:N)}\right) = \frac{ p(\vec{y})}{p\left(X^{(1:N)}\right)}\prod_{i=1}^{N} \frac{p\left(\vec{x}^{(i)}\right)}{p(\vec{y})}p\left(\vec{y}|\vec{x}^{(i)}\right).
\end{equation}
Equation (\ref{eq:combination_with_normalization}) contains many difficult to calculate terms that do not depend on $\vec{y}$, namely $p\left(X^{(1:N)}\right)$ and $p\left(\vec{x}^{(i)}\right)$. To get rid of these, we can express the equation as a score:

\noindent \begin{multline}
\label{eq:combinatorial_theorem}
    \nabla_{\vec{y}}\log p\left(\vec{y}|X^{(1:N)}\right) = \nabla_{\vec{y}}\log p(\vec{y}) + \\ \sum_{i=1}^{N} \left( \nabla_{\vec{y}}\log p\left(\vec{y}|\vec{x}^{(i)}\right) - \nabla_{\vec{y}}\log p(\vec{y}) \right). 
\end{multline}
It is thus possible to estimate the posterior score using only the individual conditional scores and the prior score. Of remark are two properties of equation (\ref{eq:combinatorial_theorem}). Firstly, if there is only 1 signal ($N=1$), then it reads as a simple identity. Secondly, the term inside the summation, $\left( \nabla_{\vec{y}}\log p\left(\vec{y}|\vec{x}^{(i)}\right) - \nabla_{\vec{y}}\log p(\vec{y}) \right)$, can be interpreted as: what additional information do we learn about $\vec{y}$ from $\vec{x}^{(i)}$, that was not already in the prior? 

 \subsubsection{Score-based diffusion modeling} \noindent
The posterior score calculated using equation (\ref{eq:combinatorial_theorem}) can be used to draw samples from $p\left(\vec{y}|X^{(1:N)}\right)$ by leveraging score-based diffusion modeling \cite{conditional_diffusion_2}. This type of generative model has garnered a lot of attention recently, due to its ease of training, stability, and high-fidelity outputs. We make use of the unifying framework proposed by Karras \textit{et al.} \cite{Karras_EDM}, who demonstrate that many different diffusion models are special cases of their framework. This enables easier expansion of the proposed FSDM model to other diffusion models in future work. Here, we briefly introduce score-based diffusion models as detailed by Karras \textit{et al.} \cite{Karras_EDM}.

In score-based diffusion models, one always starts from an easy to sample latent distribution which is subsequently transformed into the desired data distribution. Following the literature \cite{conditional_diffusion_2, Karras_EDM, Vincent_denoising_score}, a Gaussian latent distribution was chosen as starting point, $\vec{y}_0\sim\mathcal{N}(0,\sigma_{max}^2I)$. The factorized posterior score was then used to progressively move towards more likely outputs in M discrete steps, until we approximated $\vec{y}_M \sim p_{data}$. This `movement' is described by the following probability flow ordinary differential equation (ODE):
\begin{equation}
    \label{eq:ODE}
    d\vec{y} = -\dot{\sigma}(t)\sigma(t)\nabla_{\vec{y}}\log p\left(\vec{y}|X^{(1:N)}\right)dt,
\end{equation}
where $\sigma(t)$ is known as the diffusion noise schedule\footnote{\noindent Please note that the diffusion noise is a mathematical construct needed to train diffusion models, it has no relation to physical sensor noise that might be present in measurements. To avoid confusion, we will refer to the noise needed to train a diffusion model as `diffusion noise'.}, which defines the diffusion noise level at time $t$, and $\dot{\sigma}(t)$ is the first derivative with respect to $t$. Note that the probability flow ODE works both in forward as well as reverse time. To link the $M$ discrete steps to the continuous time $t$, we use the time schedule as proposed by Karras \textit{et al.} \cite{Karras_EDM}:
\begin{equation}
    t_m = \begin{cases} \left(\sigma_{max}^{1/\rho} + \frac{m}{M-1} (\sigma_{min}^{1/\rho} -\sigma_{max}^{1/\rho})\right)^\rho & \text{if $m \neq M$}\\
    ~0 & \text{if $m=M$}
\end{cases}
\end{equation}
We empirically choose $\sigma_{min} = 0.0001$, $\sigma_{max}=40$, $\rho=7$, and $M=32$. The diffusion noise schedule is simply chosen to be $\sigma(t) = t$, $\dot{\sigma}(t) = 1$.

 \subsubsection{Learning the individual conditional scores} \noindent
In order to make use of equations (\ref{eq:combinatorial_theorem}) and (\ref{eq:ODE}), estimates of the individual conditional scores are needed. To that end, we employ denoising score-matching \cite{Vincent_denoising_score}. In this framework, the scores are only estimated at chosen time steps $t$ and corresponding diffusion noise levels $\sigma(t)$ by using a denoising function in conjunction with Tweedie's approximation \cite{efron2011tweedie}. This approximation makes it much simpler to train score functions, as one only needs to train on a straight-forward denoising objective, which is easier to estimate than generic scores. Following Tweedie's approximation \cite{efron2011tweedie}, the difussion noise-level specific score estimates can be written as:
\begin{align}
\label{eq:conditional_Tweedies}
    \nabla_{\vec{y}}\log p\left(\vec{y}|\vec{x}^{(i)}\right) & \approx s_{\theta^{(i)}}\left(\vec{y}, \vec{x}^{(i)}, \sigma \right) \notag \\
    & \approx \left(D_{\theta^{(i)}}\left(\vec{y}, \vec{x}^{(i)}, \sigma \right)-\vec{y}\right)/\sigma^2,
\end{align} 
where $D_{\theta^{(i)}}$ is a denoising function implemented with neural network parameterized by $\theta^{(i)}$ and specific to the signal with index $i$. To train the denoising networks, $D_{\theta^{(i)}}$, we require a dataset of ground truth hypnograms, $\vec{y}$, with simultaneously acquired signals, $X^{(1:N)}$. We call the dataset distribution as:
\begin{equation}
    X^{(1:N)},\vec{y} \sim p_{data}^{(1:N)}. 
\end{equation}
In practice, not all signals will be measured in each recording. For example, in one recording in the dataset, sensors A-B-C might have been used, while for another recording, sensors B-C-D might have been used. To overcome this issue, each denoising network is trained only on the subset of recordings where its sensor was applied. We will denote these subsets as:
\begin{align}
    x^{(i)},y &\sim p_{data}^{(i)}
\end{align}

Since $\vec{y}$ is categorical, we train the denoising networks with the expected cross entropy loss $J$ over a range of $\sigma$ values:
\begin{align}
    \label{eq:Loss}
    J_i = - &\mathbb{E}_{x^{(i)},y \sim p_{data}^{(i)}}\mathbb{E}_{\vec{n}\sim \mathcal{N}(0,\sigma^2I)} \mathbb{E}_\sigma [ \notag \\
    & \vec{y} \log\left(D_{\theta_i}\left(\vec{y} + \vec{n}, \vec{x}^{(i)}, \sigma \right)\right) ].
\end{align}
Throughout the training process, the diffusion noise level $\sigma$ is drawn randomly using a log-normal distribution: $ln(\sigma)\sim\mathcal{N}(0.2,1.4^2)$. This biases the denoiser to minimize the loss for medium levels of diffusion noise, i.e. the diffusion noise level in the middle of the sampling  trajectory.

In summary, the individual conditional scores are approximated using denoising neural networks. These denoising neural networks are trained on the subset of recordings where their corresponding signals were measured. The loss function is the cross entropy loss, where the input to the denoiser is not only the measured signal $\vec{x}$, but also a noisy version of the ground truth hypnogram $\vec{y} + \vec{n}$ as well as the diffusion noise level $\sigma$.

\begin{figure*}
    \centering
    \includegraphics[width=0.9\linewidth, trim={0 0 0 0}]{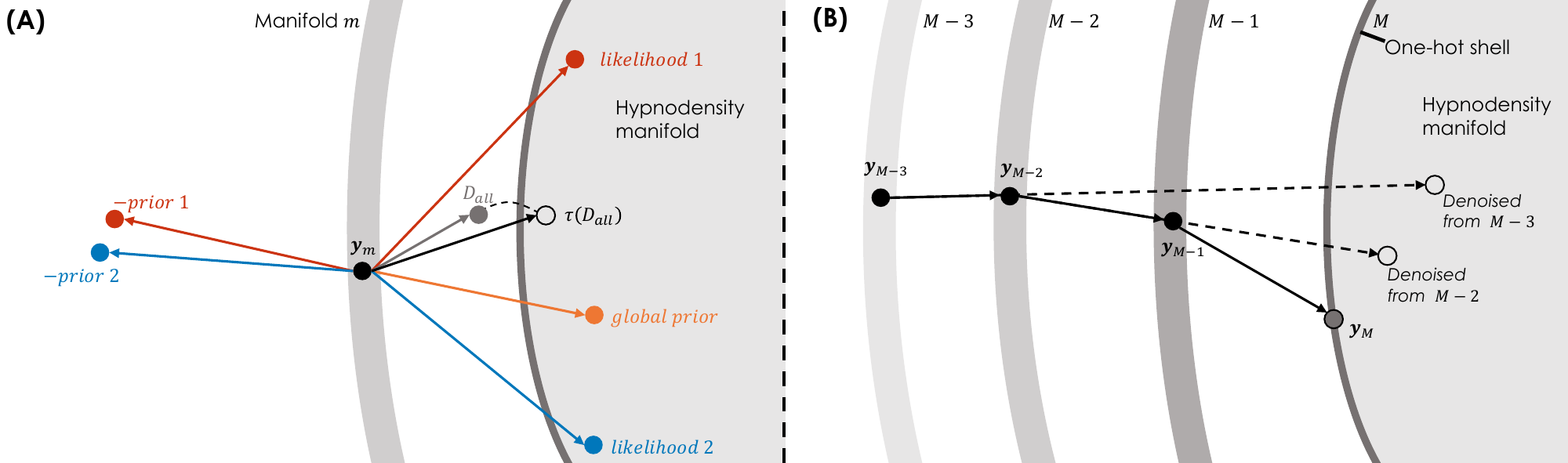}
    \caption{Visualization of the sampling process for an FSDM model. \textbf{(A)} From a current point $\vec{y}_m$ we estimate two likelihoods, two priors, and one global prior. Combining them all leads to a denoised estimate outside the hypnodensity manifold, which is corrected using a projection step $\tau()$. \textbf{(B)} Evolution of a sample over the last three time-steps. The end-estimate progressively moves from the hypnodensity manifold to the one-hot shell.}
    \label{fig:sampling_process}
    \vspace{-1em}
\end{figure*}


 \subsubsection{Learning the prior scores} \noindent
Next to the individual conditional scores, we also need to estimate the prior scores. The prior score shows up two separate times in equation (\ref{eq:combinatorial_theorem}). The first time as a `global' prior, since it is counted at the start, and the second time as an 'individual' prior, since it is subtracted from an individual likelihood score:
\begin{multline}
    \nabla_{\vec{y}}\log p\left(\vec{y}|X^{(1:N)}\right) = \underbrace{\nabla_{\vec{y}}\log p(\vec{y})}_{\text{global prior}} + \\ \sum_{i=1}^{N} \left( \underbrace{\nabla_{\vec{y}}\log p\left(\vec{y}|\vec{x}^{(i)}\right)}_{\text{individual likelihood}} - \underbrace{\nabla_{\vec{y}}\log p(\vec{y})}_{\text{individual prior}} \right). 
\end{multline}

As we discussed in the previous section, the dataset used for training does not have all signals available for all recordings. This has some implications for the prior score estimation. Specifically, if an individual likelihood score carries no new information it is desirable for the individual prior to exactly cancel it out, such that:
\begin{equation}
    \underbrace{\nabla_{\vec{y}}\log p\left(\vec{y}|\vec{x}^{(i)}\right)}_{\text{individual likelihood}} = \underbrace{\nabla_{\vec{y}}\log p(\vec{y})}_{\text{individual prior}} \text{ if } \mathcal{I}(\vec{y};\vec{x}^{(i)})=0,
\end{equation}
Where $\mathcal{I}(.;.)$ denotes mutual information. In order to achieve this behavior, we estimate each individual prior in a similar vein as equation (\ref{eq:conditional_Tweedies}) as:
\begin{align}
\label{eq:individual_prior_Tweedies}
    \underbrace{\nabla_{\vec{y}}\log p(\vec{y})}_{\text{individual prior}} & \approx s_{\theta^{(i)}}\left(\vec{y}, \vec{0}, \sigma \right) \approx \frac{D_{\theta^{(i)}}\left(\vec{y}, \vec{0}, \sigma \right)-\vec{y}}{\sigma^2},
\end{align} 
where we replaced the input signal $\vec{x}^{(i)}$ using a vector of zeroes. Additionally, we also supplement the training by augmenting the loss as specified in equation (\ref{eq:Loss}). With probability $p_{\text{augment}} = 0.5$ we partially replace the input signal $\vec{x}^{(i)}$ with some zeroes, and with probability $p_{\text{zero}} = 0.1$ we completely replace it with zeroes. This ensures that we can use each signal specific denoising network both as a conditional likelihood score estimator, as well as an individual prior estimator. The values of 0.1 and 0.5 were chosen before training and not optimized in any way.

Contrary to the individual priors, the intuition behind the global prior is that it should be trained on the largest set of possible hypnograms. To that end we train one separate global prior on the entire training dataset, since it is not constrained by sensor availability. The global prior is equal to:
\begin{align}
\label{eq:global_prior_Tweedies}
    \underbrace{\nabla_{\vec{y}}\log p(\vec{y})}_{\text{global prior}} & \approx s_{\theta^{(0)}}\left(\vec{y}, \vec{0}, \sigma \right) \approx \frac{D_{\theta^{(0)}}\left(\vec{y}, \vec{0}, \sigma \right)-\vec{y}}{\sigma^2},
\end{align} 
Where $D_{\theta^{(0)}}$ is trained using the following loss:
\begin{align}
\label{eq:loss_prior}
    J_0 = - &\mathbb{E}_{\vec{0},y \sim p_{data}^{(0)}}\mathbb{E}_{\vec{n}\sim \mathcal{N}(0,\sigma^2I)} \mathbb{E}_\sigma [ \notag \\
    & \vec{y} \log\left(D_{\theta^{(i)}}\left(\vec{y} + \vec{n}, \vec{0}, \sigma \right)\right) ],
\end{align}
where $\vec{0},y \sim p_{data}^{(0)}$ covers the entire dataset. In summary, the training of the prior networks are special cases of the conditional likelihood networks where the input signals $\vec{x}$ are (partially) set to zero. Pseudo-code of the training loop is given in Supplement K.

 \subsubsection{Sampling from an FSDM} \noindent
We can now rewrite equation (\ref{eq:combinatorial_theorem}) to use the estimated scores from (\ref{eq:conditional_Tweedies}), (\ref{eq:individual_prior_Tweedies}), and (\ref{eq:global_prior_Tweedies}):
\begin{multline}
\label{eq:combination_denoisers}
    D_{all}\left(\vec{y}, X^{(1:N)}, \sigma \right) = D_{\theta^{(0)}}\left(\vec{y}, \vec{0}, \sigma \right) + \\ \lambda \sum_{i=1}^{N} \left( D_{\theta^{(i)}}\left(\vec{y}, \vec{x}^{(i)}, \sigma \right) - D_{\theta^{(i)}}\left(\vec{y}, \vec{0}, \sigma \right) \right),
\end{multline}
where $ D_{all}$ is the combined denoising function. Additionally, we have introduced a weighting term $\lambda$ that specifies the importance of the likelihood terms with respect to the prior, which is common practice in both Bayesian inference and diffusion guidance \cite{Classifier-Diffusion-Guidance}, \cite{Classifier-Free-Diffusion-Guidance}. We empirically choose $\lambda = 1/N$, which gives rise to the desirable property that adding the same signal any number of times leads to the same posterior score. 

In practice, combining score estimates from multiple models can lead to instability in the sampling process, as the current estimate at a time-step can `fall off the manifold'. A lot of research has been done on this effect for image restoration and multiple solutions have been found \cite{conditional_diffusion_2}, \cite{pigdm_diffusion, dps_diffusion}. However, these methods assume there is some (partially) sampled data, coupling the diffusion process via a likelihood function, which is not the case for our setup as the hypnograms are completely unknown a-prior. We thus propose a new manifold projection step suited for categorical data. Since we know that on the denoised end-estimate manifold, all classes should count up to a total of probability 1, we use:
\begin{equation}    
    \tau(\vec{y}) = \vec{y} / \sum_{j=1}^{5} y_j.
\end{equation}
In other words, we re-normalize the hypnodensity to follow the hypnodensity manifold constraint.

After applying the manifold projection step, we can use the end-estimate together with Tweedie's formula to get the posterior score estimate as:
\begin{equation}
\label{eq:end_tweedies}
    \left.p\left(\vec{y}|X^{(1:N)}\right)\right\vert_\sigma \approx \left(\tau\left(D_{all}\left(\vec{y}, X^{(1:N)}, \sigma \right)\right) - \vec{y}\right)/\sigma^2.
\end{equation}

Following \cite{Karras_EDM} we employ Heunn's second order method to solve the ODE from equation (\ref{eq:ODE}) using the posterior score estimate of equation (\ref{eq:end_tweedies}). The pseudo-code for this sampling algorithm is shown in Supplement K. A visual overview of the FSDM rule is shown in Fig.\ref{fig:sampling_process}.

To generate the final hypnograms that are shown to the user and which were compared to the ground truth, we sample 64 times from the FSDM algorithm. This results in 64 different realizations of the posterior distribution $\vec{y}\sim p_{\theta}(\vec{y}|X)$, i.e., hypnograms that are likely given the input data. The end result is then calculated as the majority vote of these hypnograms:
\begin{equation}
\label{eq:majority_vote}
    \hat{\vec{h}} = \text{arg max }\mathbb{E}_{\vec{y}\sim p_{\theta}(\vec{y}|X)} \left[ \vec{y}\right],
\end{equation}
where the arg max is applied along the first dimension of $\vec{y}\in\mathcal{R}^{5\times E}$, resulting in a hypnogram with categorical sleep stages $\hat{\vec{h}}\in [W,N1,N2,N3,R]^E$. Additionally, the 64 samples are used to separately calculate the overnight sleep statistics, similar to previous work \cite{U-Flow-JBHI}. The final value for each overnight statistic for each recording is then calculated as the median of the individual realizations:
\begin{equation}
\label{eq:stat_generative}
    stat = \mathbb{\text{median}}_{\vec{y}\sim p_{\theta}(\vec{y}|X)} \left[ f_{stat}(\vec{y}) \right],
\end{equation}
where $f_{stat}$ refers to the function that calculates the overnight statistic of interest from a hypnogram, e.g. total sleep time or wake afer sleep onset. 

 \tocless\subsection{Information} \noindent
We hypothesize that the factorized combination rule from equations (\ref{eq:combinatorial_theorem}) and (\ref{eq:combination_denoisers}) allows for the evaluation of how much each individual measurement source contributes to the end-result, possibly constituting a novel interpretability metric. This can be seen through the lens of `information' as defined by Caticha 2011: \emph{``Information is what forces a change of rational beliefs''} \cite{entropic_inference}. In our case, the `rational belief' can be interpreted as the prior score, whereas the amount of change is expressed as the difference between the likelihood score and the prior score.

There are different domains and distance functions that we could use to compare the likelihood and the prior in order to express the amount of information gain. We here choose to follow recent literature on hypnodensity, which proposes the use of the cosine distance between the two vectors of class probabilities at each epoch \cite{soft_label}, \cite{ Hypnodensity_Anderer}.

To express the amount of information, we calculate the expected cosine distance between likelihood and prior over the entire sampling trajectory:
\begin{align}
    \vec{b}_i = \mathbb{E}_{\vec{y}_0} \left[ \frac{1}{M} \sum_{m=1}^{M} \right. & \left( \text{cos dist.} \left( D_{\theta^{(i)}}\left(\vec{y}_m, \vec{x}^{(i)}, \sigma(t_m) \right), \right. \right.\notag \\ 
    & \left. \left. \left. D_{\theta^{(i)}}\left(\vec{y}_m, \vec{0}\phantom{^{(i}}, \sigma(t_m) \right) \right) \right) \right],
\end{align}
where we take the expectation over different initial states of the sampling process $\vec{y}_0 \sim \mathcal{N}(0;\sigma(t_0)^2I)$. Additionally, $\vec{b}_i$ is the information for the signal with index $i$, and `cos dist.' is the cosine distance between the two hypnodensities as estimated by the prior and likelihood denoising functions. The resulting information will be of similar length as the hypnogram and takes values between 0 and 1, i.e. $\vec{b}_i\in[0,1]^{E}$. An information of 1 means that the likelihood and prior completely disagreed over the entire sampling process, and a 0 means that they always agreed, in which case one could just as easily not measured the signal at all.

\begin{table}[hbt!]
\centering
\caption{Demographic parameters for the two datasets. `\#' refers to number, and `std.' refers to the standard deviation.}
\label{tab:demographic_data}
\begin{tabular}{l|lr|rrrr}
 &
  Parameter &
   &
  Total &
  Train &
  Val &
  Test \\ \hline \hline
 &
  \cellcolor {lightgray}Recordings &
  \cellcolor {lightgray}{[}\#{]} &
  \cellcolor {lightgray}1851 &
  \cellcolor {lightgray}1281 &
  \cellcolor {lightgray}97 &
  \cellcolor {lightgray}473 \\
 &
   &
  {[}\#{]} &
  710 &
  499 &
  30 &
  181 \\
 &
  \multirow{-2}{*}{Female} &
  {[}\%{]} &
  38.4 &
  39.0 &
  30.9 &
  38.3 \\
 &
  \cellcolor {lightgray} &
  \cellcolor {lightgray}{[}mean{]} &
  \cellcolor {lightgray}51.0 &
  \cellcolor {lightgray}50.7 &
  \cellcolor {lightgray}52.5 &
  \cellcolor {lightgray}51.5 \\
 &
  \multirow{-2}{*}{\cellcolor {lightgray}Age} &
  \cellcolor {lightgray}{[}std.{]} &
  \cellcolor {lightgray}15.7 &
  \cellcolor {lightgray}16.1 &
  \cellcolor {lightgray}15.7 &
  \cellcolor {lightgray}14.7 \\
 &
   &
  {[}mean{]} &
  25.9 &
  25.8 &
  25.9 &
  26.4 \\
 &
  \multirow{-2}{*}{BMI} &
  {[}std.{]} &
  8.2 &
  8.2 &
  8.9 &
  8 \\
 &
  \cellcolor {lightgray} &
  \cellcolor {lightgray}{[}\#{]} &
  \cellcolor {lightgray}241 &
  \cellcolor {lightgray}163 &
  \cellcolor {lightgray}12 &
  \cellcolor {lightgray}66 \\
\multirow{-9}{*}{\rotatebox[origin=c]{90}{SOMNIA \cite{Somnia}}} &
  \multirow{-2}{*}{\cellcolor {lightgray}PAP usage} &
  \cellcolor {lightgray}{[}\%{]} &
  \cellcolor {lightgray}13.0 &
  \cellcolor {lightgray}12.7 &
  \cellcolor {lightgray}12.4 &
  \cellcolor {lightgray}14.0 \\ \hline
 &
  Recordings &
  {[}\#{]} &
  96 &
  66 &
  3 &
  27 \\
 &
  \cellcolor {lightgray} &
  \cellcolor {lightgray}{[}\#{]} &
  \cellcolor {lightgray}60 &
  \cellcolor {lightgray}44 &
  \cellcolor {lightgray}2 &
  \cellcolor {lightgray}14 \\
 &
  \multirow{-2}{*}{\cellcolor {lightgray}Female} &
  \cellcolor {lightgray}{[}\%{]} &
  \cellcolor {lightgray}62.5 &
  \cellcolor {lightgray}66.7 &
  \cellcolor {lightgray}66.7 &
  \cellcolor {lightgray}51.9 \\
 &
   &
  {[}mean{]} &
  36.0 &
  35.9 &
  33.7 &
  36.5 \\
 &
  \multirow{-2}{*}{Age} &
  {[}std.{]} &
  13.5 &
  13.3 &
  12.7 &
  14.1 \\
 &
  \cellcolor {lightgray} &
  \cellcolor {lightgray}{[}mean{]} &
  \cellcolor {lightgray}24.3 &
  \cellcolor {lightgray}24.0 &
  \cellcolor {lightgray}23.7 &
  \cellcolor {lightgray}25.2 \\
 \multirow{-7}{*}{\rotatebox[origin=c]{90}{HealthBed \cite{HealthBed}}}
 &
  \multirow{-2}{*}{\cellcolor {lightgray}BMI} &
  \cellcolor {lightgray}{[}std.{]} &
  \cellcolor {lightgray}3.2 &
  \cellcolor {lightgray}2.9 &
  \cellcolor {lightgray}3.8 &
  \cellcolor {lightgray}3.7 \\
\end{tabular}
\vspace{-1em}
\end{table}
\begin{table}[hbt!]
\centering
\vspace{-0.5em}
\caption{Primary sleep disorder diagnoses over the three splits. Note that many subject had multiple primary sleep diagnoses.}
\label{tab:diagnostic_data}
\begin{tabular}{l|rrrr} 
Diagnosis                                           & Total & Train & Val & Test \\
\hline \hline
\rowcolor{lightgray}
Insomnia disorders                                  & 613  & 418 & 29 & 166 \\
 
Obstructive sleep apnea                             & 1037 & 698 & 59 & 280 \\
\rowcolor{lightgray}
Central sleep apnea                                 & 42   & 26  & 2  & 14  \\
 
Treatment emergent-                                 & \multirow{2}{*}{6}    & \multirow{2}{*}{6}   & \multirow{2}{*}{0}  & \multirow{2}{*}{0}   \\
central sleep apnea \\
\rowcolor{lightgray}
Hypoventilation                                     & 8    & 6   & 0  & 2   \\
 
Narcolepsy                                          & 31   & 21  & 0  & 10  \\
\rowcolor{lightgray}
Other hypersomnolence-                              & & & & \\
\rowcolor{lightgray} disorders & \multirow{-2}{*}{54}   & \multirow{-2}{*}{40}  & \multirow{-2}{*}{3}  & \multirow{-2}{*}{11}  \\
 
Insufficient sleep syndrome                         & 66   & 52  & 3  & 11  \\
\rowcolor{lightgray}
Circadian rythm disorder                            & 46   & 34  & 4  & 8   \\
 
NREM Parasomnias                                    & 115  & 85  & 5  & 25  \\
\rowcolor{lightgray}
REM sleep behavior disorder                         & 122  & 84  & 8  & 30  \\
 
Other REM Parasomnias                               & 55   & 47  & 2  & 6   \\

\rowcolor{lightgray}
Other Parasomnias                                   & 45   & 31  & 3  & 11  \\
Restless legs syndrome and/or  & & & & \\                        
Periodic limb movement disorder & \multirow{-2}{*}{268}   & \multirow{-2}{*}{198}  & \multirow{-2}{*}{10}  & \multirow{-2}{*}{60}  \\

\rowcolor{lightgray}
Other movement disorders                            & 58   & 37  & 3  & 18  \\
 
Other sleep disorders                               & 16   & 11  & 2  & 3   \\
\rowcolor{lightgray}
No primary sleep diagnosis-  & & & & \\                        
\rowcolor{lightgray} and/or normal   variants & \multirow{-2}{*}{99}   & \multirow{-2}{*}{76}  & \multirow{-2}{*}{5}  & \multirow{-2}{*}{18}  \\
 \hline
Healthy                                             & 96   & 66  & 3  & 27 
\end{tabular} 
\end{table}
 \tocless\subsection{The SOMNIA and HealthBed datasets} \noindent
To evaluate the proposed method on a large set of signals we leveraged the Sleep and OSA Monitoring with Non-Invasive Applications (SOMNIA) dataset \cite{Somnia} and the HealthBed dataset \cite{HealthBed}. Both datasets comprise overnight polysomnographic recordings captured at Sleep Medicine Center Kempenhaeghe. The SOMNIA data comes from a diverse clinical population (1851 recordings), while the HealthBed dataset comes from healthy participants without sleep disorders (96 recordings). We included all recordings obtained in the period between 2017-01-01 and 2023-10-10 and we excluded pediatric recordings. No other exclusion criteria were applied. Note that we also included recordings where a subject made use of a PAP device during the night from which we measured PAP flow, this was the case for 241 recordings in the SOMNIA dataset.

We randomly split the recordings into 1347 train, 100 validation, and 500 hold-out test recordings. We chose to use a larger percentage of recordings in the hold-out test set than is typical to increase the diversity of pathologies in the hold-out test set. Table \ref{tab:demographic_data} shows the demographic data of each of the three sets. The distribution of primary sleep disorder categories is shown in Table \ref{tab:diagnostic_data}. Note that many subjects had multiple sleep disorder diagnoses, so the columns add up to more than the total number of recordings in each split. Supplement G provides a full breakdown per specific diagnosis.

Both the SOMNIA and HealthBed studies adhered to the guidelines of the Declaration of Helsinki, Good Clinical Practice, and current legal requirements. Both studies were reviewed by the  Maxima Medical Center medical ethical committee, Veldhoven, the Netherlands (SOMNIA: N16.074, HealthBed: NL63360.015.17), by the institutional review board of Philips (Internal Committee on Biomedical Experiments, Eindhoven, The Netherlands) and by the institutional review board of Sleep Medicine Centre Kempenhaeghe (Heeze, The Netherlands). The data analysis protocol for this study  was approved by the institutional review board of Sleep Medicine Centre Kempenhaeghe (Heeze, The Netherlands), reported under CSG.KH.2023.24 and approved on 13 November 2023.

 \tocless\subsection{Signal extraction} \noindent
A total of 36 signals were extracted from the recordings. These were grouped into 18 clusters of similar type, which we will refer to as a `signal group'. For example, the signals F3-M2, F4-M1, C3-M3, F4-M1, O1-M2, and O2-M1 were all grouped into the EEG type, and the signals from the abdominal belt and the thoracic belt were grouped into a common Respiratory Inductance Plethysmography (RIP) belt type. A full breakdown of each signal and its filtering settings can be found in Supplemental information J. The PAP flow signal was measured during overnight recordings where the subject used a PAP device, which was the case for 241 recordings, see Table \ref{tab:demographic_data}. Subjects were allowed to bring their personal PAP device, resulting in a large variety of types and manufacturers. The types of PAP included in the study were continuous PAP (CPAP) (208), auto-adjusting PAP (APAP) (23), adaptive servo ventilation (ASV) (2), and Bi-Level PAP (8).

The SOMNIA set reflects the patient population seen at the Kempenhaeghe clinic, where CPAP is the most commonly applied PAP modality.  ASV and Bi-Level PAP have specific clinical indications, often in more complex clinical cases, and the low count may be partly due to the fact that these patients were less often invited to participate in the SOMNIA study. Typically, PAP devices allow for some type pressure or flow readout. To homogenize this readout between the different devices, a common third-party sensor was attached to the breathing tube of the PAP device, called the pneumo flow (Braebon, Canada). This readout is used in the study as the `PAP flow' signal.

Recordings of all sensors were made using a dedicated recording and amplification system (Grael PSG, Compumedics, USA). The individual sensors used for each signal are described in detail in the SOMNIA protocol paper \cite{Somnia}. We applied a simple preprocessing pipeline described in Supplement J. No further filtering or other preprocessing was applied. The scoring of the ground truth sleep stages was done by expert scorers using the most recent AASM criteria at the time of each recording, which were performed in the period between 2017-01-01 and 2023-10-10.

 \tocless\subsection{Neural network architecture} \noindent
Our method is agnostic to the exact architecture used for each denoising neural network. In this manuscript, we leverage the DDPM++ model as implemented by Karras \textit{et al.} \cite{Karras_EDM}, and modified to work on 1D timeseries in our previous work on EOG-driven sleep staging \cite{EOG_Gorp}. In this section, we highlight the most important modifications made to the original DDPM++ model, see Supplement I for a complete overview of the model implementation.

Firstly, The model takes as input not only the current sample point $\vec{y}_{m-1}$ and diffusion noise level $\sigma(t_m)$, but also a conditioning created from the measured signal as $\vec{c}^{(i)} = \text{enc}(\vec{x}^{(i)})$. This conditioning is of the same size as $\vec{y}_{m-1}$ and appended channel-wise to it as input to the DDPM++ model. The conditioning networks, $\text{enc}()$, are implemented using the ResNet blocks that make up the backbone of the DDPM++ model. Using 5 levels, with 2 ResNet blocks per level, and a final strided convolution, these conditioning networks compress the input signals $\vec{x}^{(i)}\in\mathcal{R}^{1792\cdot30\cdot128}$ to conditioning vectors $\vec{c}^{(i)} \in \mathcal{R}^{1792\times16}$, i.e. a length of 1792 with 16 channels. Note that the ResNet blocks typically use a timestep embedding, but these are not added to the epoch encoder. This speeds up the sampling process, as the epoch encoder only needs to be run once, and its output context vector can be cached.

Secondly, the DDPM++ model makes use of self-attention, to which we added positional encoding using sine-cosine embedding, creating a transformer encoder layer. The transformer architecture enables the model to learn the temporal relations within and between the signals and hypnograms.

Thirdly, we adapted DDPM++ to work on 1D time-series, using 1D convolutions of kernel size 7 with 32 channels. We used 4 resolution levels with a down-sampling stride of 4. We applied a transformer layer at all resolution levels.

 \tocless\subsection{Metrics} \noindent
To evaluate the performance of the FSDM framework, we use the average accuracy, Cohen's kappa, and the macro F1 score. Additional metrics, such as the unweighted average recall (UAR), are provided in supplemental material B. We calculate each metric per recording, to subsequently average them over all recordings. This is done to ensure that shorter recordings count equally to the final result as longer recordings. 

To compare the overnight statistics from FSDM to the ground truth overnight statistics, we use Bland-Altman plots. These plots visualize the agreement by plotting the differences against their averages. Additionally, the bias and 95\% limits of agreement are calculated to identify any systematic offset or significant disagreement.

To quantitatively evaluate our novel interpretability metric based on information gain, we calculate the average information gain for each single-sensor setup and compare it to the classification performance on the hold-out test set. We assess the correlation between performance and information gain using Pearson's correlation. Additionally, we evaluate the effects of noise and missing data by removing signal segments, replacing them with zeros, or adding Gaussian noise at different signal-to-noise ratios (SNRs). The previously fitted linear relationship is applied to this new experiment to test its robustness.

\begin{table*}[hbt!]
\centering
\caption{Average hold-out test set metrics of all models for full five-class sleep staging. Colored boxes indicate signal combinations, with the indentations signifying which signals were part of the combination. Acc: accuracy, MF1: macro F1 score, PPG: photoplethysmography, SCM: sternocleidomastoid, FDS: flexor digitorum superficialis, IHR: instantaneous heart rate, IBR: instantaneous breathing rate.} \vspace{-0.5em}
\label{tab:main_results}

\begin{tabular}{ccc}

\begin{tabular}{l|ccc} 
 & \multicolumn{3}{c}{\textbf{Five-class staging}} \\
\textbf{Signal (combinations)} & \textbf{Acc [\%]} & \textbf{Kappa [-]} & \textbf{MF1 [-]} \\ 
\hline 
\hline 
\cellcolor{orange}\textcolor{white}{                       All PSG electrodes}  & \cellcolor{orange}\textcolor{white}{85.9}  & \cellcolor{orange}\textcolor{white}{0.793}  & \cellcolor{orange}\textcolor{white}{0.813} \\ 
\cellcolor{green}\textcolor{white}{\textcolor{orange}{$\bm{\vert}$} \quad All odd EEG electrodes}  & \cellcolor{green}\textcolor{white}{85.5}  & \cellcolor{green}\textcolor{white}{0.789}  & \cellcolor{green}\textcolor{white}{0.810} \\ 
\cellcolor{lightgray}\textcolor{black}{\textcolor{orange}{$\bm{\vert}$} \quad \textcolor{green}{$\bm{\vert}$} \quad F3-M2 (EEG)}  & \cellcolor{lightgray}\textcolor{black}{85.6}  & \cellcolor{lightgray}\textcolor{black}{0.791}  & \cellcolor{lightgray}\textcolor{black}{0.810} \\ 
\cellcolor{white}\textcolor{black}{\textcolor{orange}{$\bm{\vert}$} \quad \textcolor{green}{$\bm{\vert}$} \quad C3-M2 (EEG)}  & \cellcolor{white}\textcolor{black}{85.3}  & \cellcolor{white}\textcolor{black}{0.787}  & \cellcolor{white}\textcolor{black}{0.809} \\ 
\cellcolor{lightgray}\textcolor{black}{\textcolor{orange}{$\bm{\vert}$} \quad \textcolor{green}{$\bm{\vert}$} \quad O1-M2 (EEG)}  & \cellcolor{lightgray}\textcolor{black}{83.4}  & \cellcolor{lightgray}\textcolor{black}{0.760}  & \cellcolor{lightgray}\textcolor{black}{0.785} \\ 
\cellcolor{red}\textcolor{white}{\textcolor{orange}{$\bm{\vert}$} \quad Recommended PSG electrodes}  & \cellcolor{red}\textcolor{white}{86.3}  & \cellcolor{red}\textcolor{white}{0.799}  & \cellcolor{red}\textcolor{white}{0.819} \\ 
\cellcolor{blue}\textcolor{white}{\textcolor{orange}{$\bm{\vert}$} \quad \textcolor{red}{$\bm{\vert}$} \quad All even EEG electrodes}  & \cellcolor{blue}\textcolor{white}{85.8}  & \cellcolor{blue}\textcolor{white}{0.794}  & \cellcolor{blue}\textcolor{white}{0.813} \\ 
\cellcolor{white}\textcolor{black}{\textcolor{orange}{$\bm{\vert}$} \quad \textcolor{red}{$\bm{\vert}$} \quad \textcolor{blue}{$\bm{\vert}$} \quad F4-M1 (EEG)}  & \cellcolor{white}\textcolor{black}{85.5}  & \cellcolor{white}\textcolor{black}{0.790}  & \cellcolor{white}\textcolor{black}{0.808} \\ 
\cellcolor{lightgray}\textcolor{black}{\textcolor{orange}{$\bm{\vert}$} \quad \textcolor{red}{$\bm{\vert}$} \quad \textcolor{blue}{$\bm{\vert}$} \quad C4-M1 (EEG)}  & \cellcolor{lightgray}\textcolor{black}{85.5}  & \cellcolor{lightgray}\textcolor{black}{0.791}  & \cellcolor{lightgray}\textcolor{black}{0.810} \\ 
\cellcolor{white}\textcolor{black}{\textcolor{orange}{$\bm{\vert}$} \quad \textcolor{red}{$\bm{\vert}$} \quad \textcolor{blue}{$\bm{\vert}$} \quad O2-M1 (EEG)}  & \cellcolor{white}\textcolor{black}{83.8}  & \cellcolor{white}\textcolor{black}{0.764}  & \cellcolor{white}\textcolor{black}{0.790} \\ 
\cellcolor{lightgray}\textcolor{black}{\textcolor{orange}{$\bm{\vert}$} \quad \textcolor{red}{$\bm{\vert}$} \quad E2-M2 (EOG)}  & \cellcolor{lightgray}\textcolor{black}{85.0}  & \cellcolor{lightgray}\textcolor{black}{0.784}  & \cellcolor{lightgray}\textcolor{black}{0.806} \\ 
\cellcolor{white}\textcolor{black}{\textcolor{orange}{$\bm{\vert}$} \quad \textcolor{red}{$\bm{\vert}$} \quad E1-M2 (EOG)}  & \cellcolor{white}\textcolor{black}{84.3}  & \cellcolor{white}\textcolor{black}{0.776}  & \cellcolor{white}\textcolor{black}{0.801} \\ 
\cellcolor{lightgray}\textcolor{black}{\textcolor{orange}{$\bm{\vert}$} \quad \textcolor{red}{$\bm{\vert}$} \quad Chin1-ChinZ (EMG)}  & \cellcolor{lightgray}\textcolor{black}{74.9}  & \cellcolor{lightgray}\textcolor{black}{0.630}  & \cellcolor{lightgray}\textcolor{black}{0.676} \\ 
\cellcolor{white}\textcolor{black}{\textcolor{orange}{$\bm{\vert}$} \quad Chin2-ChinZ (EMG)}  & \cellcolor{white}\textcolor{black}{74.9}  & \cellcolor{white}\textcolor{black}{0.631}  & \cellcolor{white}\textcolor{black}{0.677} \\ 
\cellcolor{lightgray}\textcolor{black}{\textcolor{orange}{$\bm{\vert}$} \quad Chin1-Chin2 (EMG)}  & \cellcolor{lightgray}\textcolor{black}{74.5}  & \cellcolor{lightgray}\textcolor{black}{0.624}  & \cellcolor{lightgray}\textcolor{black}{0.672} \\ 
\hline \hline 
\cellcolor{pink}\textcolor{white}{  HSAT expanded}  & \cellcolor{pink}\textcolor{white}{79.0}  & \cellcolor{pink}\textcolor{white}{0.697}  & \cellcolor{pink}\textcolor{white}{0.729} \\ 
\cellcolor{lightblue}\textcolor{white}{\textcolor{pink}{$\bm{\vert}$} \quad HSAT reduced}  & \cellcolor{lightblue}\textcolor{white}{78.3}  & \cellcolor{lightblue}\textcolor{white}{0.686}  & \cellcolor{lightblue}\textcolor{white}{0.715} \\ 
\cellcolor{white}\textcolor{black}{\textcolor{pink}{$\bm{\vert}$} \quad \textcolor{lightblue}{$\bm{\vert}$} \quad Nasal cannula}  & \cellcolor{white}\textcolor{black}{76.5}  & \cellcolor{white}\textcolor{black}{0.661}  & \cellcolor{white}\textcolor{black}{0.695} \\ 
\cellcolor{lightgray}\textcolor{black}{\textcolor{pink}{$\bm{\vert}$} \quad \textcolor{lightblue}{$\bm{\vert}$} \quad Finger PPG}  & \cellcolor{lightgray}\textcolor{black}{75.1}  & \cellcolor{lightgray}\textcolor{black}{0.640}  & \cellcolor{lightgray}\textcolor{black}{0.681} \\ 
\cellcolor{white}\textcolor{black}{\textcolor{pink}{$\bm{\vert}$} \quad Thoracic belt}  & \cellcolor{white}\textcolor{black}{76.3}  & \cellcolor{white}\textcolor{black}{0.657}  & \cellcolor{white}\textcolor{black}{0.708} \\ 
\hline \hline 
\cellcolor{pink}\textcolor{white}{  HSAT expanded}  & \cellcolor{pink}\textcolor{white}{78.5}  & \cellcolor{pink}\textcolor{white}{0.687}  & \cellcolor{pink}\textcolor{white}{0.722} \\ 
\cellcolor{lightblue}\textcolor{white}{\textcolor{pink}{$\bm{\vert}$} \quad HSAT reduced}  & \cellcolor{lightblue}\textcolor{white}{77.7}  & \cellcolor{lightblue}\textcolor{white}{0.674}  & \cellcolor{lightblue}\textcolor{white}{0.707} \\ 
\cellcolor{lightgray}\textcolor{black}{\textcolor{pink}{$\bm{\vert}$} \quad \textcolor{lightblue}{$\bm{\vert}$} \quad thermistor}  & \cellcolor{lightgray}\textcolor{black}{72.9}  & \cellcolor{lightgray}\textcolor{black}{0.603}  & \cellcolor{lightgray}\textcolor{black}{0.652} \\ 
\cellcolor{white}\textcolor{black}{\textcolor{pink}{$\bm{\vert}$} \quad \textcolor{lightblue}{$\bm{\vert}$} \quad ECG}  & \cellcolor{white}\textcolor{black}{76.9}  & \cellcolor{white}\textcolor{black}{0.669}  & \cellcolor{white}\textcolor{black}{0.706} \\ 
\cellcolor{lightgray}\textcolor{black}{\textcolor{pink}{$\bm{\vert}$} \quad Thoracic belt}  & \cellcolor{lightgray}\textcolor{black}{76.3}  & \cellcolor{lightgray}\textcolor{black}{0.657}  & \cellcolor{lightgray}\textcolor{black}{0.708} \\ 
\hline \hline 
\cellcolor{pink}\textcolor{white}{  HSAT expanded}  & \cellcolor{pink}\textcolor{white}{78.2}  & \cellcolor{pink}\textcolor{white}{0.678}  & \cellcolor{pink}\textcolor{white}{0.704} \\ 
\cellcolor{lightblue}\textcolor{white}{\textcolor{pink}{$\bm{\vert}$} \quad HSAT reduced}  & \cellcolor{lightblue}\textcolor{white}{76.4}  & \cellcolor{lightblue}\textcolor{white}{0.652}  & \cellcolor{lightblue}\textcolor{white}{0.684} \\ 
\cellcolor{white}\textcolor{black}{\textcolor{pink}{$\bm{\vert}$} \quad \textcolor{lightblue}{$\bm{\vert}$} \quad PAP flow}  & \cellcolor{white}\textcolor{black}{69.5}  & \cellcolor{white}\textcolor{black}{0.562}  & \cellcolor{white}\textcolor{black}{0.610} \\ 
\cellcolor{lightgray}\textcolor{black}{\textcolor{pink}{$\bm{\vert}$} \quad \textcolor{lightblue}{$\bm{\vert}$} \quad Finger PPG}  & \cellcolor{lightgray}\textcolor{black}{75.1}  & \cellcolor{lightgray}\textcolor{black}{0.640}  & \cellcolor{lightgray}\textcolor{black}{0.681} \\ 
\cellcolor{white}\textcolor{black}{\textcolor{pink}{$\bm{\vert}$} \quad Thoracic belt}  & \cellcolor{white}\textcolor{black}{76.3}  & \cellcolor{white}\textcolor{black}{0.657}  & \cellcolor{white}\textcolor{black}{0.708} \\ 
\hline \multicolumn{4}{c}{\phantom{a}}
\end{tabular} 

&
 
& 

\begin{tabular}{l|ccc} 
 & \multicolumn{3}{c}{\textbf{Five-class staging}} \\
\textbf{Signal (combinations)} & \textbf{Acc [\%]} & \textbf{Kappa [-]} & \textbf{MF1 [-]} \\ 
\hline 
\hline 
\cellcolor{lightblue}\textcolor{white}{          HSAT reduced}  & \cellcolor{lightblue}\textcolor{white}{77.6}  & \cellcolor{lightblue}\textcolor{white}{0.676}  & \cellcolor{lightblue}\textcolor{white}{0.710} \\ 
\cellcolor{white}\textcolor{black}{\textcolor{lightblue}{$\bm{\vert}$} \quad Nasal cannula}  & \cellcolor{white}\textcolor{black}{76.5}  & \cellcolor{white}\textcolor{black}{0.661}  & \cellcolor{white}\textcolor{black}{0.695} \\ 
\cellcolor{lightgray}\textcolor{black}{\textcolor{lightblue}{$\bm{\vert}$} \quad IHR from finger PPG}  & \cellcolor{lightgray}\textcolor{black}{71.8}  & \cellcolor{lightgray}\textcolor{black}{0.597}  & \cellcolor{lightgray}\textcolor{black}{0.651} \\ 
\hline \hline 
\cellcolor{lightblue}\textcolor{white}{          HSAT reduced}  & \cellcolor{lightblue}\textcolor{white}{74.9}  & \cellcolor{lightblue}\textcolor{white}{0.626}  & \cellcolor{lightblue}\textcolor{white}{0.653} \\ 
\cellcolor{lightgray}\textcolor{black}{\textcolor{lightblue}{$\bm{\vert}$} \quad IBR from PAP flow}  & \cellcolor{lightgray}\textcolor{black}{69.2}  & \cellcolor{lightgray}\textcolor{black}{0.538}  & \cellcolor{lightgray}\textcolor{black}{0.581} \\ 
\cellcolor{white}\textcolor{black}{\textcolor{lightblue}{$\bm{\vert}$} \quad IHR from finger PPG}  & \cellcolor{white}\textcolor{black}{71.8}  & \cellcolor{white}\textcolor{black}{0.597}  & \cellcolor{white}\textcolor{black}{0.651} \\ 
\hline \hline 
\cellcolor{red}\textcolor{white}{Left Leg and SCM}  & \cellcolor{red}\textcolor{white}{71.0}  & \cellcolor{red}\textcolor{white}{0.575}  & \cellcolor{red}\textcolor{white}{0.621} \\ 
\cellcolor{white}\textcolor{black}{\textcolor{red}{$\bm{\vert}$} \quad Left Leg (EMG)}  & \cellcolor{white}\textcolor{black}{66.9}  & \cellcolor{white}\textcolor{black}{0.526}  & \cellcolor{white}\textcolor{black}{0.586} \\ 
\cellcolor{lightgray}\textcolor{black}{\textcolor{red}{$\bm{\vert}$} \quad Left SCM (EMG)}  & \cellcolor{lightgray}\textcolor{black}{66.2}  & \cellcolor{lightgray}\textcolor{black}{0.502}  & \cellcolor{lightgray}\textcolor{black}{0.575} \\ 
\hline \hline 
\cellcolor{red}\textcolor{white}{Right Leg and SCM}  & \cellcolor{red}\textcolor{white}{70.3}  & \cellcolor{red}\textcolor{white}{0.565}  & \cellcolor{red}\textcolor{white}{0.616} \\ 
\cellcolor{lightgray}\textcolor{black}{\textcolor{red}{$\bm{\vert}$} \quad Right Leg (EMG)}  & \cellcolor{lightgray}\textcolor{black}{66.7}  & \cellcolor{lightgray}\textcolor{black}{0.523}  & \cellcolor{lightgray}\textcolor{black}{0.582} \\ 
\cellcolor{white}\textcolor{black}{\textcolor{red}{$\bm{\vert}$} \quad Right SCM (EMG)}  & \cellcolor{white}\textcolor{black}{67.0}  & \cellcolor{white}\textcolor{black}{0.517}  & \cellcolor{white}\textcolor{black}{0.572} \\ 
\hline \hline 
\cellcolor{red}\textcolor{white}{Left Leg and FDS}  & \cellcolor{red}\textcolor{white}{67.4}  & \cellcolor{red}\textcolor{white}{0.532}  & \cellcolor{red}\textcolor{white}{0.599} \\ 
\cellcolor{white}\textcolor{black}{\textcolor{red}{$\bm{\vert}$} \quad Left Leg (EMG)}  & \cellcolor{white}\textcolor{black}{66.9}  & \cellcolor{white}\textcolor{black}{0.526}  & \cellcolor{white}\textcolor{black}{0.586} \\ 
\cellcolor{lightgray}\textcolor{black}{\textcolor{red}{$\bm{\vert}$} \quad Left FDS (EMG)}  & \cellcolor{lightgray}\textcolor{black}{63.7}  & \cellcolor{lightgray}\textcolor{black}{0.482}  & \cellcolor{lightgray}\textcolor{black}{0.566} \\ 
\hline \hline 
\cellcolor{red}\textcolor{white}{Right Leg and FDS}  & \cellcolor{red}\textcolor{white}{68.0}  & \cellcolor{red}\textcolor{white}{0.540}  & \cellcolor{red}\textcolor{white}{0.604} \\ 
\cellcolor{lightgray}\textcolor{black}{\textcolor{red}{$\bm{\vert}$} \quad Right Leg (EMG)}  & \cellcolor{lightgray}\textcolor{black}{66.7}  & \cellcolor{lightgray}\textcolor{black}{0.523}  & \cellcolor{lightgray}\textcolor{black}{0.582} \\ 
\cellcolor{white}\textcolor{black}{\textcolor{red}{$\bm{\vert}$} \quad Right FDS (EMG)}  & \cellcolor{white}\textcolor{black}{63.2}  & \cellcolor{white}\textcolor{black}{0.476}  & \cellcolor{white}\textcolor{black}{0.560} \\ 
\hline \hline 
\cellcolor{lightgray}\textcolor{black}{Abdominal belt}  & \cellcolor{lightgray}\textcolor{black}{76.4}  & \cellcolor{lightgray}\textcolor{black}{0.661}  & \cellcolor{lightgray}\textcolor{black}{0.712} \\ 
\cellcolor{white}\textcolor{black}{Snore microphone}  & \cellcolor{white}\textcolor{black}{72.1}  & \cellcolor{white}\textcolor{black}{0.597}  & \cellcolor{white}\textcolor{black}{0.654} \\ 
\cellcolor{lightgray}\textcolor{black}{IHR from ECG}  & \cellcolor{lightgray}\textcolor{black}{70.1}  & \cellcolor{lightgray}\textcolor{black}{0.571}  & \cellcolor{lightgray}\textcolor{black}{0.634} \\ 
\cellcolor{white}\textcolor{black}{IBR from RIP thorax}  & \cellcolor{white}\textcolor{black}{70.0}  & \cellcolor{white}\textcolor{black}{0.564}  & \cellcolor{white}\textcolor{black}{0.606} \\ 
\cellcolor{lightgray}\textcolor{black}{IBR from RIP abdomen}  & \cellcolor{lightgray}\textcolor{black}{69.9}  & \cellcolor{lightgray}\textcolor{black}{0.563}  & \cellcolor{lightgray}\textcolor{black}{0.605} \\ 
\cellcolor{white}\textcolor{black}{SpO2}  & \cellcolor{white}\textcolor{black}{68.7}  & \cellcolor{white}\textcolor{black}{0.542}  & \cellcolor{white}\textcolor{black}{0.592} \\ 
\cellcolor{lightgray}\textcolor{black}{IBR from nasal cannula}  & \cellcolor{lightgray}\textcolor{black}{66.9}  & \cellcolor{lightgray}\textcolor{black}{0.521}  & \cellcolor{lightgray}\textcolor{black}{0.577} \\ 
\cellcolor{white}\textcolor{black}{IBR from Thermistor}  & \cellcolor{white}\textcolor{black}{65.4}  & \cellcolor{white}\textcolor{black}{0.497}  & \cellcolor{white}\textcolor{black}{0.557} \\ 
\cellcolor{lightgray}\textcolor{black}{Suprasternal notch}  & \cellcolor{lightgray}\textcolor{black}{63.1}  & \cellcolor{lightgray}\textcolor{black}{0.464}  & \cellcolor{lightgray}\textcolor{black}{0.546} \\ 
\cellcolor{white}\textcolor{black}{IBR esophageal pressure}  & \cellcolor{white}\textcolor{black}{59.0}  & \cellcolor{white}\textcolor{black}{0.419}  & \cellcolor{white}\textcolor{black}{0.504} \\ 
\cellcolor{lightgray}\textcolor{black}{IBR Suprasternal notch}  & \cellcolor{lightgray}\textcolor{black}{58.5}  & \cellcolor{lightgray}\textcolor{black}{0.394}  & \cellcolor{lightgray}\textcolor{black}{0.489} \\ 
\cellcolor{white}\textcolor{black}{Esophageal pressure}  & \cellcolor{white}\textcolor{black}{55.5}  & \cellcolor{white}\textcolor{black}{0.373}  & \cellcolor{white}\textcolor{black}{0.485} \\ \hline 
\end{tabular} 

\end{tabular}

\vspace{-1em}
\end{table*}
 \tocless\subsection{Additional comparison on Sleep-EDF expanded} \noindent
To test the proposed FSDM model on a different dataset and compare it to literature, we applied our model to the EDF expanded dataset from 2018 \cite{sleepedf}, \cite{goldberger2000physiobank}. We used the Sleep Cassette set, which consists of 78 healthy sleepers who underwent 2 consecutive nights of recording. Each recording lasts up to 20 hours, and subjects wore a modified Walkman-like cassette-tape recorder that measured several signals at sampling rates of 100 Hz or 1Hz. To align with our methodology, we only used the signals measured at 100 Hz, which are: EEG Fpz-Cz, EEG Pz-Oz, and Horizontal EOG. Scoring was performed by experts following the Rechtschaffen \& Kales rules \cite{RandK} which were adapted to the AASM standard by merging stages S3 and S4 into N3. Furthermore, epochs scored as 'Movement Time' were disregarded when calculating performance metrics. Following the literature, the recordings were cropped to $\pm$30 minutes around the subject's sleep \cite{SleepTransformer}.

We tested two approaches on this dataset. First, we tested direct application, where we directly used the model weights trained on SOMNIA and HealthBed on Sleep-EDF Expanded. Second, we experimented with training our model from scratch. Following the literature, we applied 10-fold cross-validation, constantly leaving 7-8 subjects out for hold-out testing. We also left one other fold out as a validation set and trained on the remaining 8 folds. We trained models on each of the three signals: EEG Fpz-Cz, EEG Pz-Oz, and Horizontal EOG.
\begin{figure*}
    \centering
    \includegraphics[width=0.9\linewidth, trim={0 1em 0 0}]{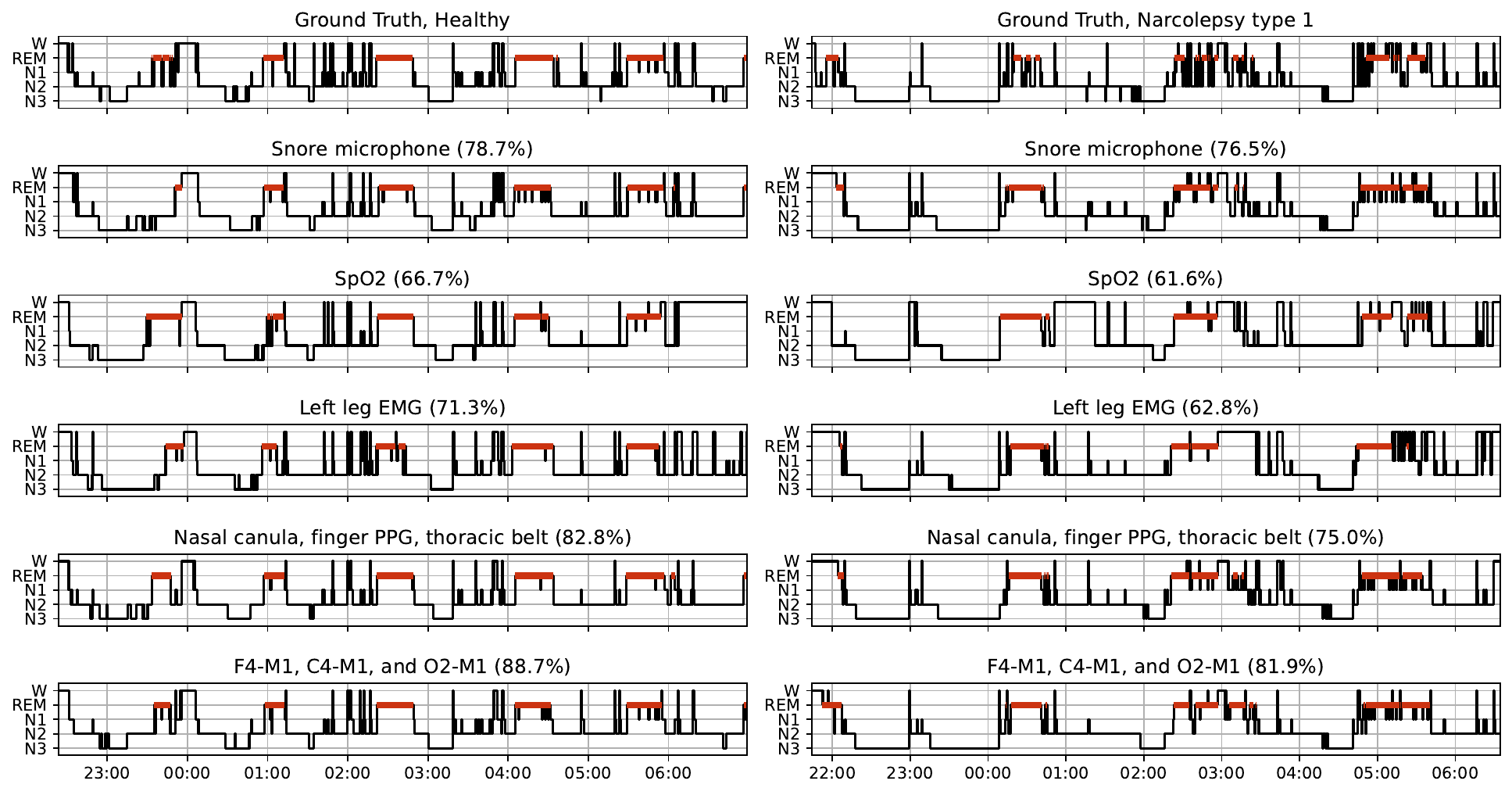}
    \caption{Qualitative examples of using five different signal combinations on a healthy subject (left column), and on a subject with narcolepsy type 1 (right column). The 5-class accuracy on each recording is listed between brackets. The red bars denote REM sleep.}
    \label{fig:qualitative}
\end{figure*}
 \tocless\section{Results} \noindent
We evaluated the proposed method in terms of agreement with the human scored hypnogram, testing each of the 36 signals individually and when used in combinations. Table \ref{tab:main_results} shows the resulting 5-class performance metrics on the test set. Supplement B shows additional results for other setups and metrics like 4-class sleep staging and UAR.

The resulting metrics for the single-channel EEG and EOG models are the highest out of all the evaluated signals, e.g. achieving accuracies between 83.4\% and 85.6\%, indicating that any of these signals on their own enables high quality sleep staging. Using these neurological signals together further improves performance, with the highest accuracy of 86.3\% achieved when using the combination of signals for sleep staging recommended by the AASM \cite{AASM}.

Table \ref{tab:main_results} also shows the results for several signal combinations typically available with home sleep apnea tests (HSATs) \cite{AASM}. These were split into reduced HSATs, which combine a cardiac signal with a respiratory flow signal, and expanded HSATs, which add a respiratory effort signal as well. We observe that using these signals in combination leads to better sleep staging performance as compared to using each of these signals on their own. Furthermore, the expanded HSATs improve upon their respective reduced sets, indicating that the respiratory effort signal further supplements the information available in the cardiac and respiratory flow signals. Similar results were found when combining two EMG signals, where the combination results in higher accuracy than each of the signals individually. Please see Supplement N for a comparison of our HSAT perfromance to those reported in literature. 

Qualitative results for a subset of the signal (combinations) is shown in Fig. \ref{fig:qualitative}, which shows the results for a healthy subject and one with narcolepsy type 1 who displayed a sleep onset REM period (SOREMP), which is one of the diagnostic criteria of that sleep disorder \cite{ICSD3_latest}. These examples were selected to illustrate performance on both healthy sleep and a sleep disorder that manifests in the hypnogram. Table \ref{tab:main_results} and Fig. \ref{fig:qualitative} show that unconventional signals can also be leveraged to perform sleep staging, albeit at a lower accuracy. For example, the SpO2 signal reaches an average 5-class accuracy of 68.7\% and reasonably captures the overall shape of the hypnograms in Fig. \ref{fig:qualitative}. It however misses the SOREMP in the subject with narcolepsy of Fig. \ref{fig:qualitative}. The more conventional signal combinations fare much better in this regard, with especially the EEG combination clearly detecting the SOREMP. 

\begin{figure*}
    \centering
    \begin{tabular}{c}
    \includegraphics[width=0.85\linewidth, trim={0 0.6cm 0 0}]{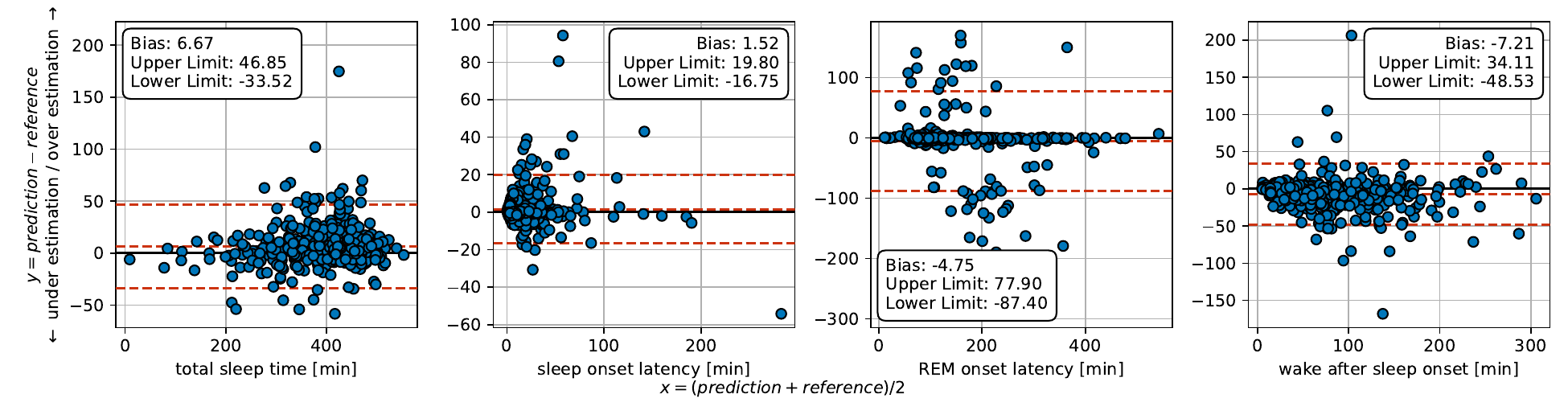} \\
    \includegraphics[width=0.85\linewidth, trim={0 0.3cm 0 0}]{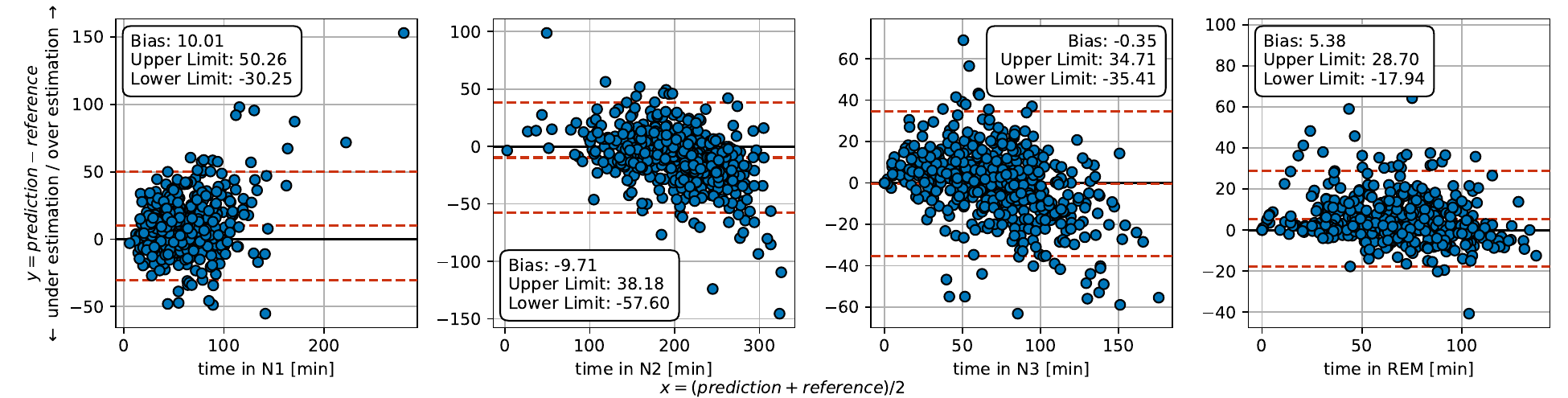} \\
    \end{tabular}
    \caption{Bland-Altman plots for the overnight sleep statistics as predicted by the recommended PSG setup over all recordings in the hold-out test set. The limits of agreement are given at the 95\% confidence interval. A positive y value indicates an overestimation by our model with respect to the gold-standard, while a negative value indicates an underestimation.}    
    \label{fig:bland-altman-all}
    \vspace{-1em}
\end{figure*}

\begin{figure*}
    \centering
    \includegraphics[width=0.85\linewidth, trim={0 0.3cm 0 0}]{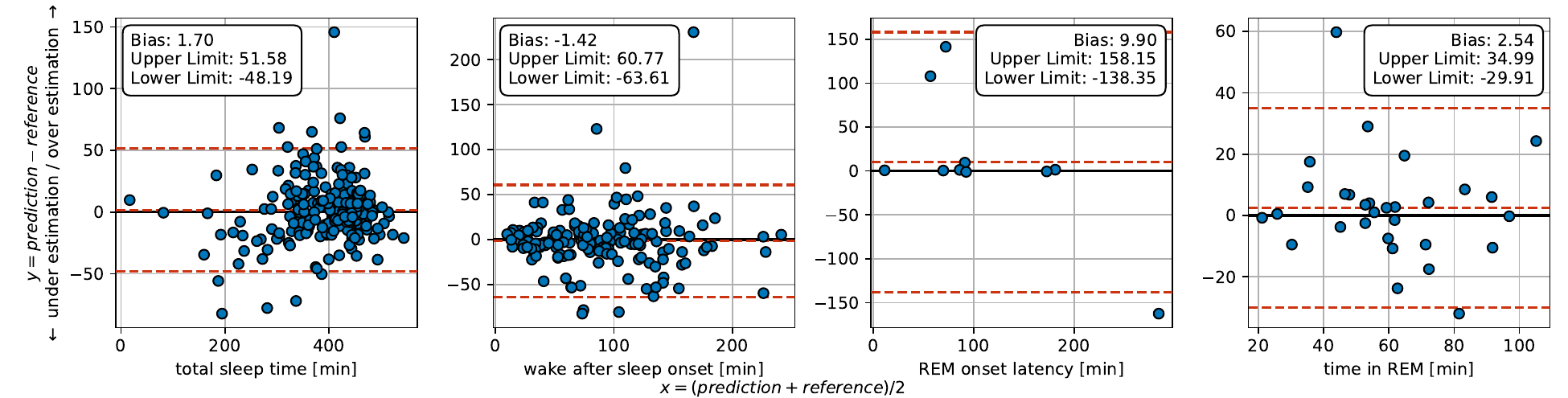}
    \caption{Bland-Altman plots for four combinations of sleep statistics, disorders, and input signals. Limits of agreement are at the 95\% confidence interval. From left to right: total sleep time for OSA (HSAT: Nasal Cannula + finger PPG + Thoracic Belt), WASO [min] for insomnia (finger PPG), REM onset latency [min] for narcolepsy (recommended PSG setup), and time in REM [min] for RBD (single channel EEG: F4-M1).}   
    \label{fig:bland-altman-diagnoses}
    \vspace{-1em}
\end{figure*}

Fig. \ref{fig:bland-altman-all} and Fig. \ref{fig:bland-altman-diagnoses} show two evaluations of the estimation of the overnight sleep statistics. Fig. \ref{fig:bland-altman-all} displays the Bland-Altman plots that result form estimating the overnight sleep statistics over all recordings in the hold-out test set using the AASM recommended PSG set-up as input. Fig. \ref{fig:bland-altman-diagnoses} shows four Bland-Altman plots for specific combinations of overnight sleep statistic and input signal(s) evaluated only on the subjects in the hold-out test set with a certain disorder, which were chosen to highlight relevant use-cases in sleep medicine and research. For example, our method is able to measure the total sleep time for subjects with obstructive sleep apnea (OSA) using an HSAT, a parameter that is typically not captured by that measurement setup. We also show wake after sleep onset (WASO) estimation for insomnia subjects with a finger PPG, REM onset latency for narcolepsy subjects using a PSG set-up, and time in REM for subjects with REM sleep behavior disorder (RBD) measured with a single-channel EEG.

\begin{figure*}
    \centering
    \begin{tabular}{cc}
    \includegraphics[width=0.4\linewidth, trim={0 0 0 0}]{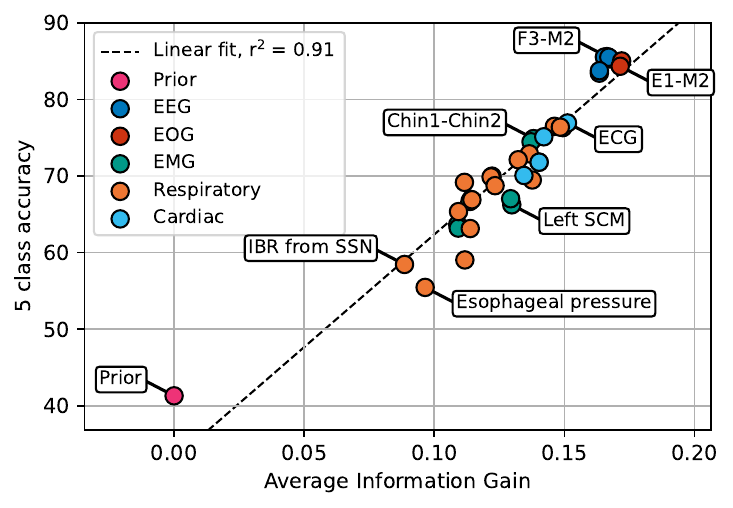} &
    \includegraphics[width=0.4\linewidth, trim={0 0 0 0}]{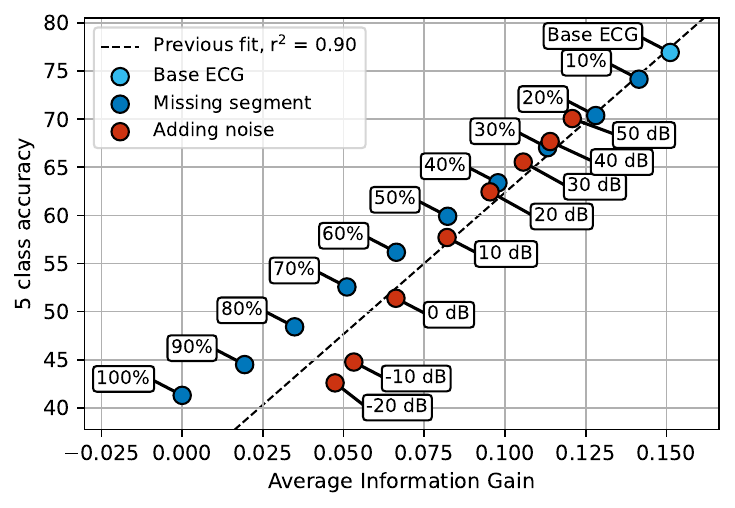} \\
    \end{tabular}
    \caption{Average information gain, i.e. the average difference between likelihood and prior terms, versus 5-class accuracy. Left: Linear correlation between information gain per sensor and accuracy, with highlighted sensor positions. Right: Impact of reducing ECG signal quality (removing segments or adding noise) on accuracy and information gain, with text boxes showing the percentage of recording removed or SNR of added noise. The linear fit from the left plot still provides a good fit. \textit{SCM}: sternocleidomastoid, \textit{IBR}: instantaneous breathing rate, \textit{SSN}: suprasternal notch.}
    \label{fig:quantitative-information}
\end{figure*}

In general, the proposed method displays low bias and high agreement. Some outliers can be observed in the estimation of sleep onset latency and REM onset latency. These happen due to the all-or-nothing nature in the estimation of these statistics. For example, if a period of REM is missed by the method the REM onset latency will be postponed by an entire sleep cycle leading to over-estimations in the order of 100 minutes, while if the method (falsely) detects a period of REM sleep an entire cycle earlier, this leads to under-estimations.  These outliers do not mean that the assembled hypnogram is not reliable, as the other overnight statistics are not impacted by this effect and epoch-to-epoch agreement with the ground truth is high.

The correlation between the average information gain for each single-sensor setup and the classification accuracy on the hold-out test set is shown in Fig. \ref{fig:quantitative-information} on the left. A strong linear correlation was found between these two metrics (Pearson's correlation coefficient of 0.91). The effects of noise at different SNRs and the replacing of segments with zeroes for the ECG signal are shown in Fig. \ref{fig:quantitative-information} on the right. The linear fit as found for the sensors still held for the retrospective addition of these artifacts, as direct application of this line achieves a Pearson's correlation coefficient of 0.90. Supplement E shows this experiment with Cohen's kappa and the F3-M2 signal.

\begin{figure*}[hbt!]
    \centering
    \includegraphics[width=0.9\linewidth, trim={0 0 0 0}]{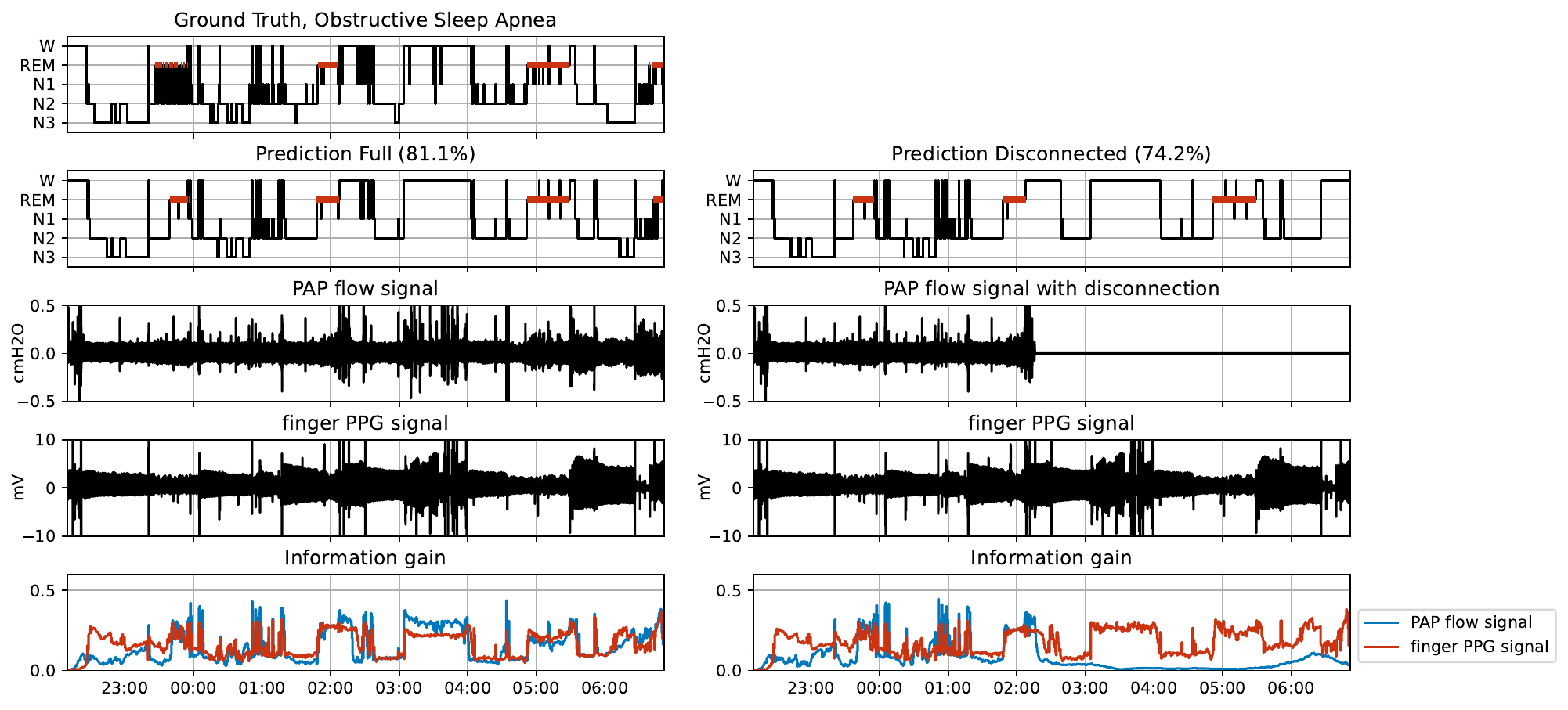}
    \vspace{-1em}
    \caption{The proposed method is robust to the disconnection of sensors. Left, output of using the PAP flow and finger PPG signals over the entire night. Right, artificially created example of what would happen if the user took off their PAP device halfway through the night at 2:15 hours. Bottom row, our novel interpretability metric in terms of per-signal information gain, calculated as the difference between each signal's likelihood and prior. It can be observed that in the disconnection case, the information gain from the PAP flow goes to zero after 2:15 hours.}
    \label{fig:disconnected_sensor}
    \vspace{-1em}
\end{figure*}

A qualitative example of the information gain metric is shown in Fig. \ref{fig:disconnected_sensor}, which displays the information gain per epoch when using PAP flow and finger PPG for a subject with obstructive sleep apnea (OSA). Another example for cannula with PPG is shown in supplemental information F. The information gain is calculated as the difference between the signal likelihood and the prior, where the prior estimates the probability of a hypnogram as compared to hypnograms seen in the training set. Samples from the prior are shown in Supplement D. \newpage

Fig. \ref{fig:disconnected_sensor} shows that the information gain is low when the prediction coincides with the prior, for example when predicting wake at the start of the recording or N2 during the night, while information gain is high when it departs from the prior, for example when predicting a long sequence of wake in the middle of the night or a period of N3 at the end of the night.

Similar to the segment of zeroes as tested and illustrated in Fig. \ref{fig:quantitative-information}, a post-hoc simulation of sensor disconnection was introduced into the recording. This result is shown in the right column of Fig. \ref{fig:disconnected_sensor}. Here, we artificially simulated the scenario where the user of a PAP device takes off their mask during the night, in this case at 2:15 hours, while keeping the PPG device connected. The model elegantly handles this situation and is still able to perform adequate sleep staging, even with such a sensor disconnection. From the information gain plots in the last row of Fig. \ref{fig:disconnected_sensor}, it can also be observed how the information gained from the PAP flow signal goes to zero in the second half of the night, except for the final awakening, as the model has learned to correlate switching off of devices with awakenings. We can observe that the sleep staging prediction does suffer in quality from the sensor disconnection, as it for example misses the very last N3 and REM periods between 6:00 and 7:00.

Table \ref{tab:edf78} shows the results on the Sleep EDF expanded dataset for direct application of the SOMNIA-trained models as well as training from scratch using 10-fold cross validation. We also compare our model to results from literature that used the same 10-fold cross validation setup. Note that there is a mismatch between the signals available in SOMNIA and those in the EDF expanded dataset. To allow for a direct application of the SOMNIA-trained models, we used the models trained on the signals that are topographically closest. 

Our model achieves highly comparable performance to that of other models proposed in the literature. Even though our model does not reach the highest performance of all proposed methods, it has the major benefit of being much more flexible and scalable due to the factorized combination rule. Table \ref{tab:edf78} also shows that training from scratch on Sleep-EDF resulted in better performance than the direct application approach, probably due to the differences in signal acquisition, scoring rules, recording equipment and especially electrode placement. Interestingly, the horizontal EOG from the Sleep-EDF dataset does not reach the same sleep staging performance as the E1-M2 and E2-M2 electrodes from the SOMNIA set, even when training from scratch. This suggests that the horizontal EOG acquired in the Sleep-EDF dataset has different characteristics and is not as informative for sleep staging as the new AASM-recommended EOG placement \cite{AASM}.
\begin{table*}[hbt!]
\centering
\caption{Results on the `EDF Expanded' dataset. We show the results for direct application of the SOMNIA trained models as well as training from scratch using 10-fold cross validation cropped to $\pm$ 30 around the subject's sleep. Results from literature are taken from their respective papers and all used simmilar 10-fold cross validation and cropping.}
\label{tab:edf78}
\begin{tabular}{ll|ccc}
& & \multicolumn{3}{c}{\textbf{Five-class staging}} \\
\textbf{Method} & \textbf{Signals} & \textbf{Acc [\%]} & \textbf{Kappa [-]} & \textbf{MF1 [-]} \\ \hline \hline
\rowcolor{lightgray} 
FSDM trained on SOMNIA $\dagger \ddagger$ & Fpz-Cz + Pz-Oz + Horizontal EOG & 81.9 & 0.746 & 0.751 \\
FSDM trained on SOMNIA $\dagger \ddagger$ & Fpz-Cz + Horizontal EOG & 79.5 & 0.715 & 0.732 \\
\rowcolor{lightgray} 
FSDM trained on SOMNIA $\dagger$ & Fpz-Cz + Pz-Oz & 81.7 & 0.741 & 0.746\\
FSDM trained on SOMNIA $\dagger$ & Fpz-Cz & 76.1 & 0.662 & 0.696\\
\rowcolor{lightgray} 
FSDM trained on SOMNIA $\dagger$ & Pz-Oz & 80.7 & 0.728 & 0.734\\
FSDM trained on SOMNIA $\ddagger$ & Horizontal EOG & 76.4 & 0.675 & 0.706\\ \hline
\rowcolor{lightgray} 
FSDM trained on EDF Expanded& Fpz-Cz + Pz-Oz + Horizontal EOG & 82.9 & 0.758 & 0.762\\
FSDM trained on EDF Expanded& Fpz-Cz + Horizontal EOG & 82.6 & 0.753 & 0.757\\
\rowcolor{lightgray} 
FSDM trained on EDF Expanded& Fpz-Cz + Pz-Oz & 82.7 & 0.755 & 0.756\\ 
FSDM trained on EDF Expanded& Fpz-Cz & 82.5 & 0.753 & 0.754\\
\rowcolor{lightgray} 
FSDM trained on EDF Expanded & Pz-Oz & 79.9 & 0.715 & 0.724\\
FSDM trained on EDF Expanded& Horizontal EOG & 79.2 & 0.704 & 0.723\\ \hline
\rowcolor{lightgray} 
Catboost  \cite{van2023not}  (2023) & Fpz-Cz + Pz-Oz + Horizontal EOG & 83.0 & 0.763 & 0.772 \\
CareSleepNet \cite{wang2024caresleepnet} (2024) & Fpz-Cz + Horizontal EOG & 85.1 & 0.789 & 0.804 \\
\rowcolor{lightgray} 
XSleepNet2 \cite{phan2021xsleepnet} (2022) & Fpz-Cz + Horizontal EOG & 84.0 & 0.778 & 0.779 \\
XSleepNet2 \cite{phan2021xsleepnet} (2022) & Fpz-Cz & 84.0 & 0.778 & 0.787 \\
\rowcolor{lightgray} 
LGSleepNet \cite{shen2023lgsleepnet} (2023) & Fpz-Cz & 82.3 & 0.75\phantom{0} & 0.760 \\
SleepTransformer \cite{SleepTransformer} (2022) & Fpz-Cz & 81.4 & 0.743 & 0.743 \\
\rowcolor{lightgray} 
TinySleepNet \cite{supratak2020tinysleepnet} (2021) & Fpz-Cz & 83.1 & 0.77\phantom{0} & 0.781 \\
DeepSleepNet-Lite \cite{fiorillo2021deepsleepnet} (2021) & Fpz-Cz & 80.3 & 0.73\phantom{0} & 0.752 \\
\rowcolor{lightgray} 
U-Time \cite{perslev_utime} (2019) & Fpz-Cz & - & - & 0.76\phantom{0} \\
SleepEEGNet \cite{mousavi2019sleepeegnet} (2019) & Fpz-Cz & 80.0 & 0.73\phantom{0} & 0.736\\
\rowcolor{lightgray} 
SleepEEGNet \cite{mousavi2019sleepeegnet} (2019) & Pz-Oz & 77.6 & 0.689 & 0.700
\\

\end{tabular} \\
\vspace{0.4em}
There is a mismatch between signal derivations present in SOMNIA and those in EDF expanded: \\
$\dagger$ Using the model trained on derivations F3-M2, F4-M1, C3-M2, C4-M1, O1-M2, O2-M1.\\
$\ddagger$ Using the model trained on derivations E1-M2, E2-M2.
\vspace{-2em}
\end{table*}
 \tocless\section{Discussion} \noindent
We introduced a deep generative model for sleep staging with arbitrary sensor input. We have shown that the model can be applied to not only standard sleep staging signals, such as the EEG, or surrogate signals, such as the finger PPG, but also to unconventional ones, such as the SpO2 signal or the Leg EMG. Additionally, by leveraging the factorized score-based diffusion rule, the model can be applied to any combination of sensor inputs using separately trained models. This results in two desirable properties. Firstly, the proposed system can be extended naturally to newly developed measurement modalities, as one only needs to train one separate network solely on this new signal, after which it can be seamlessly adopted into our framework. This makes the framework highly scalable, because the integration of a new sensor does not require the collection of recordings with all sensors present. This is a unique feature resulting from the application of diffusion modeling to Bayes’ rule, allowing it to circumvent the intractable evidence terms from equation (\ref{eq:combination_with_normalization}). Secondly, the ad-hoc combination of sensors, especially surrogate sensors, opens up sleep staging to other fields of medicine, where sleep is largely under-studied but still of vital importance. For example, allowing identification of sleep disorders from Holter ECG in the context of cardiac arrythmias or tracking sleep based on vital signs in the ICU.

The factorized score-based diffusion rule permits a natural means of expressing the information gained from each input signal by comparing its likelihood and prior terms. Following recent literature \cite{soft_label}, \cite{Hypnodensity_Anderer}, this comparison was performed using the cosine distance between the hypnodensities generated by the terms. This is different from current interpretability approaches in automatic sleep staging, which can broadly be categorized into three strategies. Gradient-based approaches that leverage the Gradient-weighted Class Activation Mapping (Grad-CAM) algorithm \cite{gradcam_sleep1}, \cite{gradcam_sleep2}, which relate how much each input element contributes to the classification output. Attention-based approaches, which characterizes which parts of the input space are used by the model for each decision, leveraging both ‘soft’ \cite{SleepTransformer} and ‘hard’ attention \cite{Huijben2024}. Lastly, SHapley Additive exPlanations (SHAP) methods, which assign an additive importance value to each input feature \cite{shap_sleep}. The information metric contrasts with the aforementioned approaches, as it is not concerned with relating the decision to each specific sample (or mini-epoch) in the input signals. Rather, it calculates for each epoch how much information each signal contributed, with information defined as a divergence from rational beliefs (the prior). The information gain was found to be strongly correlated to classification performance. Furthermore the average information gain drops when a signal becomes less ‘useful’, for example during segments of missing data, or in the presence of additive noise.

To achieve the unique advantages of the factorized score-based diffusion model, we assumed conditional independence of the signals given the hypnogram. Without this assumption, key features such as separate training of models, information gain calculation, and zero-shot inference on unseen combinations of signals would not be possible. The assumption of conditional independence is similar to that of the naive Bayesian classifier. However, while the naive Bayesian classifier is linear, the proposed FSDM model is highly non-linear due to its reliance on score-based diffusion and neural networks. In practice, the assumption of conditional independence may not always hold, for example through effects like cardio-respiratory coupling or neuro-cardiac coupling. However, introducing inter-dependence between modalities would require training the model on combinations of signals, which would negate the advantages of our approach. This holds for any type of machine learning model (neural networks as well as conventional algorithms). Despite not fully modeling such inter-dependencies, our model still achieves sleep staging performance comparable to human inter-rater agreement, empirically demonstrating that the violation of this assumption does not impact the results.

The proposed method is related to mixture of experts (MOE) models \cite{moe2},   where different parts of a network specialize in distinct tasks. In the proposed FSDM framework, we create an expert for each sleep measurement signal. However, the FSDM algorithm differs from traditional MOE models in several key aspects. Typically, MOE models are applied to a single input data type, after which a gating network determines which expert(s) to utilize for producing an output and how to combine their contributions. In contrast, we leverage diverse input types, each with its own specific expert. Additionally, no gating network is employed. Instead, we use our factorized score calculation from equation (\ref{eq:FSDM_rule}). This approach offers a unique advantage over MOEs. In MOEs, the gating network needs to be trained and, if applied to our framework, would still require examples with all signals present simultaneously. We circumvent this issue with the theoretical setup of equation (1), as we only need to train on examples of individual signals.

In literature on automatic sleep staging, human inter-rater agreement serves as an upper limit on performance since it characterizes how consistent the ground truth is to which we are comparing \cite{Uncertainty_Gorp}. The large-scale study by Rosenberg and Van Hout conducted based on the AASM inter-rater agreement program \cite{agreementRosenberg} found an average agreement of 82.6\% using the scoring behavior of over 2,500 scorers. Because the data used in the present study came from one clinic, its inter-rater agreement serves as the upper limit. An average agreement of around 86\% has been measured, based on both an internal institutional inter-rater agreement assessment and the AASM inter-rater agreement program. We verified this on the 111 recordings of the dataset where two scorings from multiple technicians were available, finding an average agreement of 85.8\%. The proposed model reaches this upper limit for inputs of single-channel EEG, single-channel EOG, or combinations that include EEG/EOG. For all other signals, it is much more difficult to ascertain whether the performance limit has been reached, as one would need to characterize exactly how much the model could be improved further (epistemic uncertainty), versus how much inherent sleep stage ambiguity is present in these signals (aleatoric uncertainty) \cite{Uncertainty_Gorp}. In particular when using surrogate signals where visual scoring by humans is not possible, these limits have not been formally established.

The present work also offers some surprising new insights on the potential of different sensors for the sleep staging task. Firstly, the relatively good performance of the snoring microphone both quantitatively and qualitatively. This microphone is placed in contact with the skin directly above the trachea and is typically only used to monitor snoring. However, we hypothesize that the vibrations caused by cardiac and respiratory activity are picked up by the microphone and enable sleep staging by our model. Secondly, the fact that it is possible to perform sleep staging using the SpO2 signal. While 5-class accuracy using the SpO2 signal is only 68.7\% (Table \ref{tab:main_results}), sleep-wake accuracy reaches 89.1\% (supplement B). Thirdly, the sleep staging performances of the EMG signals are surprisingly high, see Table \ref{tab:main_results}. Typically, these sensors are only used to detect muscle atonia during the REM stage, but we show here that they carry information about all the sleep stages. The performance of using EMG signals becomes even better when considering two EMG signals from different muscle groups, such as the leg together with the sternocleidomastoid. Lastly, our extensive analysis of various signals that can be leveraged for sleep staging on the same dataset demonstrates their relative usefulness for this task. The best results are obtained with EEG/EOG, particularly for 5-class sleep stage classification. However, other, more practical sensors can provide sufficiently accurate sleep staging performance for specific tasks. For example, total sleep time estimation can be effectively estimated with an HSAT, long-term sleep monitoring can be conducted using only a finger PPG or Holter ECG, and bruxism detection can be performed using only chin EMG.

The use of only SpO2 or single-channel EMG to perform sleep staging has not been described in the literature before. In future work, it needs to be investigated whether these findings hold across different acquisition setups. In the SOMNIA and HealthBed datasets, only minimal preprocessing is performed on the front-end of the sensors and all data is stored in high resolution, with low-pass filtering, sampling rates and quantization specifications beyond those recommended by the AASM. This is in general not the case for data measured in many sleep laboratories, where forms of data compression, quantization, filtering, and resampling, are usually applied. One hypothesis is that sleep staging based on the SpO2 and EMG signals is possible in this set, because the minimal preprocessing leaves the possibility of desirable (insofar as sleep staging is concerned) contamination of cardiac and respiratory signals. However, this hypothesis remains speculative for now and needs to be tested in future work. The desirable contamination effect has already been proven for EOG-based sleep staging, where EEG contamination can be leveraged to achieve high levels of agreement against manual scoring \cite{EOG_Gorp}, \cite{zhu2023}, and in sleep staging based on the suprasternal notch sensor, from which the cardiac and respiratory signals can be extracted \cite{cerina2023sleep}.

The use of score-based diffusion models requires some computational considerations. We trained the models on two machines (NVIDIA GeForce RTX 3080 TI, NVIDIA RTX A5000) and found that training a model takes between 30 to 60 hours and uses 7.6 GB of VRAM. It is important to note that this training time represents a one-time upfront investment and can be done in parallel for each signal modality. Once trained, the inference process is fast, as highlighted in Supplement L. Moreover, the 30-60 hours of GPU time is relatively minor compared to the substantial cost and time investment required for data collection (e.g., the dataset used in this study comprises 16,244 hours of recording). The hardware requirements, especially the VRAM usage, hamper the application of the model in edge devices. Future work focusing on reducing the computational cost could consider hyper-parameter optimization, neural network pruning, consistency models \cite{Consistency_Models_1}, and model distillation \cite{Flash_diffusion}. However, our current results, achieved without these methods, already demonstrate the robustness and effectiveness of our approach.

This work opens several avenues for future research. 
Firstly, the model could be expanded to cover often-used wearables and nearables by training signal-specific networks for each of them, these include  wrist-worn reflective PPG, under-the-mattress sensors, microphones placed near the bed, and video. Care must be taken to synchronize these devices to the PSG from which the ground truth is derived, e.g. by matching the inter-beat intervals of two cardiac signals. 
Secondly, our main dataset is not publicly available (see data availability statement for data sharing conditions). To facilitate reproducibility, we added results on the open access Sleep EDF expanded dataset. While we have evaluated the direct application of the SOMNIA models and training the models from scratch, the exploration of different transfer learning strategies can be studied in future work to obtain even better results on new datasets. 
Moreover, the model could be applied to datasets with multiple scorings available per recording to evaluate the degree of inter-rater agreement (of overnight sleep statistics) captured. This can be done using the same setup as reported in previous work \cite{U-Flow-JBHI}, where cumulative distributions of multiple scorers are compared to those created by a generative sleep staging model. 
Additionally, no extensive hyper-parameter search or tuning was performed beyond an initial verification of the chosen diffusion settings on the validation set. A comprehensive hyper-parameter optimization could be considered to potentially improve the performance of the proposed model. Care must be taken however to not introduce unintentional overfitting or test set leakage.
Furthermore, we have demonstrated that the flow in a PAP device can be leveraged to perform sleep staging, at least using CPAP or APAP devices. The number of recordings with ASV and bi-level PAP was low, so we cannot compare sleep structure of subjects using different PAP devices. Future studies could recruit larger numbers of subjects using different PAP modalities to allow for further comparisons. 
Lastly, while the FSDM framework is proposed here to factorize the score over sensor modalities, other factorizations could also be considered. For example, by factorizing over different patient populations, such as patients with intellectual disabilities or patients in an ICU, one could tailor the automatic sleep staging model to their specific characteristics. A factorization across different clinics could also be applied, which would enable federated learning where clinics only share their final model with one another and not the underlying training data. 

To conclude, we developed a deep generative model for the task of 5-class sleep staging that can use arbitrary (combinations of) sensor input. The unified framework allows for the direct comparison of different combinations of input measurements on sleep staging performance. Our proposed factorized solution is highly flexible, can be applied to a myriad of settings, and can easily be extended to new sensors while at the same time being robust to missing data. Furthermore, we proposed a novel interpretability metric based on information gain, allowing us to express at what time and by how much the model makes use of each signal for its decision. This work represents a fundamental step in the direction of a true universal sleep staging algorithm that goes beyond traditional fixed measurement set-ups and paves the way for more accessible and adaptable sleep analysis in diverse clinical populations and settings.


\section*{Acknowledgements} \noindent
We thank Luca Cerina for his help with the automatic multi-scale peak detection (AMPD) algorithm \cite{ampd}. 

This work was performed within the IMPULSE framework of the Eindhoven MedTech Innovation Center (e/MTIC, incorporating Eindhoven University of Technology, Philips Research, and Sleep Medicine Center, Kempenhaeghe Foundation), including a PPS supplement from the Dutch Ministry of Economic Affairs and Climate Policy.  P.F. reports personal fees from Philips during the conduct of the study; personal fees from Philips, outside the submitted work. S.O. received an unrestricted research grant from UCB Pharma and participated in advisory boards for Jazz Pharmaceuticals, Bioprojet, and Abbvie. All unrelated to present work.

\section*{Competing interests} \noindent
At the time of writing, H.G. and P.F. were employed and/or affiliated with Royal Philips, a commercial company and manufacturer of consumer and medical electronic devices, commercializing products in the area of sleep diagnostics and sleep therapy. Philips had no role in the study design, decision to publish, or preparation of the manuscript.

\section*{Data availability} \noindent
The SOMNIA and the HealthBed dataset are not publicly available. For academic purposes, the data can in principle be shared, subject to all applicable regulations and scope of the work. For detail see supplement A.  The Sleep EDF expanded dataset \cite{sleepedf} is available online from PhysioNet \cite{goldberger2000physiobank} at: \url{https://physionet.org/content/sleep-edfx/1.0.0/}.

\section*{Code and Supplemental information} \noindent
Supplemental information is available at: \url{https://github.com/HansvanGorp/FSDM-supplement}

The implementation of the automatic multi-scale peak detection (AMPD) algorithm \cite{ampd} used in this study is available  at: \url{https://github.com/LucaCerina/ampdLib}. Code snippets for the FSDM algorithm are available on GitHub at: \url{https://github.com/HansvanGorp/FSDM}. The other codes used in the study are proprietary, and available only with the permission of the licensors.

\renewcommand*{\bibfont}{\small}
\printbibliography

\end{document}